\documentclass[aps, prl, final, showkeys, showpacs, superscriptaddress, floatfix, nofootinbib, twocolumn, amsmath, amssymb,dvipsnames]{revtex4-2}

\usepackage{atlasphysics} 
\usepackage{graphicx}

\usepackage{paralist}
\usepackage{enumitem}
\usepackage{xcolor}
\usepackage[normalem]{ulem}
\usepackage{bm}

\usepackage[colorlinks]{hyperref}
\hypersetup{pdftitle={DarkSPHERE}, 
pdfauthor={K. Nikolopoulos and others},
urlcolor=Plum,
citecolor=WildStrawberry,
linkcolor=RoyalBlue}
\newcommand\proposalAcronym{{\sc DarkSPHERE}}

\usepackage{siunitx}
\usepackage{multirow} 
\DeclareSIUnit\parsec{pc}
\DeclareSIUnit\lightyear{ly}
\DeclareSIUnit\HOUR{hour}
\DeclareSIUnit\DAY{day}
\DeclareSIUnit\YEAR{year}
\DeclareSIUnit\microBqperkg{\si[per-mode=symbol]{\micro\becquerel\per\kilo\gram}}
\DeclareSIUnit\milliBqperkg{\si[per-mode=symbol]{\milli\becquerel\per\kilo\gram}}
\DeclareSIUnit\picogperg{\si[per-mode=symbol]{\pico\gram\per\gram}} 
\usepackage{paralist}
\usepackage{xcolor}
\usepackage{graphicx}
\graphicspath{{./Figures/}}
\usepackage{tikz}

\usepackage{siunitx}

\usepackage[caption=false]{subfig}
\usepackage{booktabs}

\begin{document}
\title{Exploring light dark matter with the DarkSPHERE spherical proportional counter electroformed underground at the Boulby Underground Laboratory}

\author{L.~Balogh}
\affiliation{Department of Mechanical and Materials Engineering, Queen’s University, Kingston, Ontario K7L 3N6, Canada}
\author{C.~Beaufort}
\affiliation{LPSC, Universit\'{e} Grenoble-Alpes, CNRS-IN2P3, Grenoble, 38026, France}
\author{M.~Chapellier}
\affiliation{Department of Physics, Engineering Physics \& Astronomy, Queen’s University, Kingston, Ontario, K7L 3N6, Canada}
\author{E. C.~Corcoran}
\affiliation{Chemistry \& Chemical Engineering Department, Royal Military College of Canada, Kingston, Ontario K7K 7B4, Canada}
\author{J.-M.~Coquillat}
\affiliation{Department of Physics, Engineering Physics \& Astronomy, Queen’s University, Kingston, Ontario, K7L 3N6, Canada}
\author{A.~Dastgheibi-Fard}
\affiliation{LPSC, Universit\'{e} Grenoble-Alpes, CNRS-IN2P3, Grenoble, 38026, France}
\author{Y.~Deng}
\affiliation{Department of Physics, University of Alberta, Edmonton, T6G 2E1, Canada}
\author{D.~Durnford}
\affiliation{Department of Physics, University of Alberta, Edmonton, T6G 2E1, Canada}
\author{C.~Garrah}
\affiliation{Department of Physics, University of Alberta, Edmonton, T6G 2E1, Canada}
\author{G.~Gerbier}
\affiliation{Department of Physics, Engineering Physics \& Astronomy, Queen’s University, Kingston, Ontario, K7L 3N6, Canada}
\author{I.~Giomataris}
\affiliation{IRFU, CEA, Universit\'{e} Paris-Saclay, F-91191 Gif-sur-Yvette, France}
\author{G.~Giroux}
\affiliation{Department of Physics, Engineering Physics \& Astronomy, Queen’s University, Kingston, Ontario, K7L 3N6, Canada}
\author{P.~Gorel}
\affiliation{SNOLAB, Lively, Ontario, P3Y 1N2, Canada}
\author{M.~Gros}
\affiliation{IRFU, CEA, Universit\'{e} Paris-Saclay, F-91191 Gif-sur-Yvette, France}
\author{P.~Gros}
\affiliation{Department of Physics, Engineering Physics \& Astronomy, Queen’s University, Kingston, Ontario, K7L 3N6, Canada}
\author{O.~Guillaudin}
\affiliation{LPSC, Universit\'{e} Grenoble-Alpes, CNRS-IN2P3, Grenoble, 38026, France}
\author{E.~W. Hoppe}
\affiliation{Pacific Northwest National Laboratory, Richland, Washington 99354, USA}
\author{I.~Katsioulas}
\affiliation{School of Physics and Astronomy, University of Birmingham, Birmingham, B15 2TT, UK}
\author{F.~Kelly}
\affiliation{Chemistry \& Chemical Engineering Department, Royal Military College of Canada, Kingston, Ontario K7K 7B4, Canada}
\author{P.~Knights}
 \email[e-mail:]{p.r.knights@bham.ac.uk}
\affiliation{School of Physics and Astronomy, University of Birmingham, Birmingham, B15 2TT, UK}
\author{P.~Lautridou}
\affiliation{SUBATECH, IMT-Atlantique/CNRS-IN2P3/Nantes University, Nantes, 44307, France}
\author{I.~Manthos}
\affiliation{School of Physics and Astronomy, University of Birmingham, Birmingham, B15 2TT, UK}
\author{R.D.~Martin}
\affiliation{Department of Physics, Engineering Physics \& Astronomy, Queen’s University, Kingston, Ontario, K7L 3N6, Canada}
\author{J.~Matthews}
\affiliation{School of Physics and Astronomy, University of Birmingham, Birmingham, B15 2TT, UK}
\author{J.-F.~Muraz}
\affiliation{LPSC, Universit\'{e} Grenoble-Alpes, CNRS-IN2P3, Grenoble, 38026, France}
\author{T.~Neep}
\affiliation{School of Physics and Astronomy, University of Birmingham, Birmingham, B15 2TT, UK}
\author{K.~Nikolopoulos}
\affiliation{School of Physics and Astronomy, University of Birmingham, Birmingham, B15 2TT, UK}
\author{P.~O'Brien}
\affiliation{Department of Physics, University of Alberta, Edmonton, T6G 2E1, Canada}
\author{M.-C.~Piro}
\affiliation{Department of Physics, University of Alberta, Edmonton, T6G 2E1, Canada}
\author{N.~Rowe}
\affiliation{Department of Physics, Engineering Physics \& Astronomy, Queen’s University, Kingston, Ontario, K7L 3N6, Canada}
\author{D.~Santos}
\affiliation{LPSC, Universit\'{e} Grenoble-Alpes, CNRS-IN2P3, Grenoble, 38026, France}
\author{G.~Savvidis}
\affiliation{Department of Physics, Engineering Physics \& Astronomy, Queen’s University, Kingston, Ontario, K7L 3N6, Canada}
\author{I.~Savvidis}
\affiliation{Aristotle University of Thessaloniki, Thessaloniki, 54124 Greece}
\author{F.~Vazquez~de~Sola~Fernandez}
\affiliation{SUBATECH, IMT-Atlantique/CNRS-IN2P3/Nantes University, Nantes, 44307, France}
\author{R.~Ward}
\affiliation{School of Physics and Astronomy, University of Birmingham, Birmingham, B15 2TT, UK}


\collaboration{NEWS-G Collaboration}

\author{E.~Banks}
\affiliation{STFC Boulby Underground Laboratory, Boulby Mine, Redcar-and-Cleveland, TS13 4UZ, UK}

\author{L.~Hamaide}
\affiliation{Department of Physics, King’s College London, Strand, London, WC2R 2LS, UK}

\author{C.~McCabe}
\affiliation{Department of Physics, King’s College London, Strand, London, WC2R 2LS, UK}

\author{K.~Mimasu}
\affiliation{Department of Physics, King’s College London, Strand, London, WC2R 2LS, UK}

\author{S.~Paling}
\affiliation{STFC Boulby Underground Laboratory, Boulby Mine, Redcar-and-Cleveland, TS13 4UZ, UK}

\begin{abstract}
We present the conceptual design and the physics potential of \proposalAcronym, a proposed 3\,m in diameter spherical proportional counter electroformed underground at the Boulby Underground Laboratory. This effort builds on the R\&D performed and experience acquired by the NEWS-G Collaboration.
\proposalAcronym\ is primarily designed to search for nuclear recoils from light dark matter in the 0.05--10\,GeV mass range. 
Electroforming the spherical shell and the implementation of a shield based on pure water ensures a background level below 0.01 dru. These, combined with the proposed helium-isobutane gas mixture, will provide sensitivity to the spin-independent nucleon cross-section of $2\times 10^{-41} \,(2\times 10^{-43})$\,cm$^2$ 
for a dark matter mass of $0.1 \,(1)$\,GeV.
The use of a hydrogen-rich gas mixture with a natural abundance of $^{13}$C  
provides
sensitivity to spin-dependent nucleon cross-sections more than two orders of magnitude below existing constraints for dark matter lighter than 1\,GeV.
The characteristics of the detector also make it suitable for searches of other dark matter signatures, including scattering of MeV-scale dark matter with electrons, and super-heavy dark matter with masses around the Planck scale that leave extended ionisation tracks in the detector.
\end{abstract}


\maketitle

\section{Introduction} 
\label{sec:intro}

The nature of Dark Matter (DM) is one of the most pressing questions in physics, as reflected by the number of ongoing and planned activities, spanning orders of magnitude in scale and complexity~\cite{Battaglieri:2017aum, Billard:2021uyg}. 
The $10$~--~$1000\;\si{\giga\eV}$ mass region has been under intense experimental scrutiny as the preferred range for Weakly Interacting Massive Particles (WIMPs)~\cite{Feng:2010gw}. 
However, the lack of conclusive evidence to-date, including from searches at colliders~\cite{Buchmueller:2017qhf, Argyropoulos:2021sav}, demands the broadening of the DM search strategy. 
Searches for lighter DM candidates below the Lee-Weinberg bound of about $2\;\si{\giga\eV}$~\cite{Lee:1977ua} are coming increasingly into focus, supported by a growing number of theory paradigms, e.g., asymmetric dark matter~\cite{Petraki:2013wwa,Zurek:2013wia}, hidden sectors~\cite{Boehm:2003hm, Fayet:2004bw, Essig:2013lka,Profumo:2015oya, Evans:2017kti}, and scenarios with a modified early-universe cosmology~\cite{Hochberg:2014dra,Kuflik:2015isi,Pappadopulo:2016pkp,DAgnolo:2019zkf}.

The large-scale liquid noble gas detectors that provide the most stringent constraints on WIMP interactions with an atomic nucleus (see e.g.,~\cite{XENON:2018voc,PandaX-4T:2021bab,LZ:2022ufs}) are not optimised for the light DM
region, below approximately $5~\si{\giga\eV}$, due to poor kinematic matching between the DM and target nucleus. There are ongoing attempts to recover sensitivity through processes such as the Migdal effect~\cite{Migdal:1939,Bernabei:2007jz,Ibe:2017yqa,Dolan:2017xbu}, which is itself under intense investigation~\cite{Nakamura:2020kex, MIGDAL2022}.
Several collaborations including CRESST~\cite{CRESST:2019jnq}, DAMIC~\cite{Castello-Mor:2020jhd}, EDELWEISS~\cite{EDELWEISS:2017lvq} and SuperCDMS~\cite{SuperCDMS:2018mne} have demonstrated sensitivity to light DM candidates by utilising smaller, cryogenic solid-state detectors with sub-keV detection thresholds.
However, scaling to larger exposures while maintaining low energy  thresholds and low background rates is challenging.

\begin{figure}[t!]
  \centering
  \includegraphics[width=0.95\linewidth]{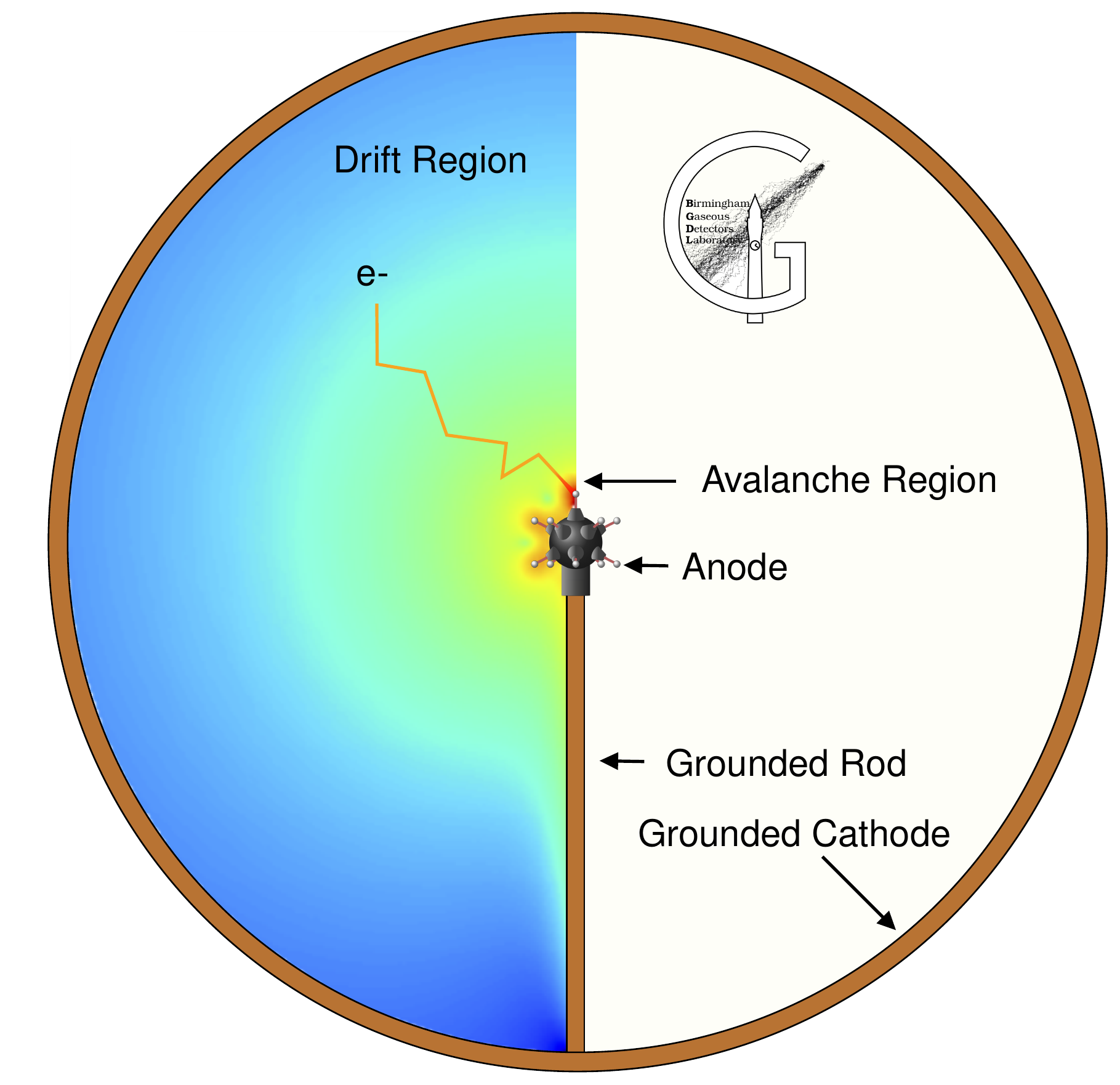}
  \caption{Schematic and principle of operation of the spherical proportional counter. Particle interactions within the gas produce ionisation electrons, which drift towards the anode at the centre of the sphere. 
  The anode shown is part of an 11-anode, multi-anode read-out system, ACHINOS. \proposalAcronym\ will operate with a multi-anode ACHINOS read-out sensor with 60-anodes. The copper used for the grounded cathode and grounded rod will be electroformed underground to minimise the background rate. 
  \label{fig:schematicSPC}}
\end{figure}

The {\it New Experiments With Spheres-Gas} (NEWS-G) Collaboration searches for light DM using spherical proportional counters~\cite{Giomataris:2003bp,Giomataris:2008ap}.
The spherical proportional counter consists of a grounded, spherical, metallic vessel filled with an appropriate gas mixture and a 
central read-out structure, with $\mathcal{O}(1\;\si{\milli\meter})$ in radius anodes,
at the centre, as depicted in Figure~\ref{fig:schematicSPC}.  
The read-out structure is supported by a grounded metallic rod, which also shields the wire used to apply a positive voltage to the anode and read out the signal.  
The electric field varies approximately as~$1/r^{2}$, dividing the gas region into a drift and an avalanche volume. Particle interactions in the gas may result in the ionisation of electrons, which subsequently drift to the anode. Within approximately 
$100\,\si{\micro\meter}$
from an anode, the electric field becomes sufficiently intense for an avalanche to occur, providing signal amplification.

Spherical proportional counters exhibit several key features that make them ideal for performing light DM searches~\cite{Gerbier:2014jwa}. 
Firstly, they have a small detector capacitance by virtue of the spherical shape, and together with the ability to operate in high gas gains, allow for the detection of nuclear recoils with sub-keV energy~\cite{Arnaud:2019nyp,Savvidis:2016wei}.  A crucial advantage is that the small detector capacitance is independent of the outer diameter $(\varnothing)$ of the detector, thus the detector can be scaled to a larger size without impacting the energy threshold. 
Secondly, the simplicity of the design, while also greatly easing detector operation, enables construction from a small number of radio-pure components. This allows for low background rates to be achieved.
Thirdly, the ability to operate the detector with different gas mixtures and pressures offers two significant benefits: 
\begin{inparaenum}[a)] 
\item the use of different atoms or molecules in the gas mixture allows kinematic DM candidate-target matching; and 
\item changes to the gas mixture and pressure provide additional handles to disentangle potential signals from unknown instrumental backgrounds. 
\end{inparaenum}
Finally, analysis of the pulse shape provides a powerful handle for background rejection and fiducialisation~\cite{Arnaud:2017bjh}.

The first DM results with a spherical proportional counter were produced using {\sc SEDINE}, a $\varnothing60\,\si{\centi\meter}$ detector operating at the Laboratoire Souterrain de Modane (LSM), France~\cite{Piquemal:2012fs}. At the time,  {\sc SEDINE} provided the best sensitivity to
the spin-independent DM-nucleon scattering cross-section
 at $0.5\,\si{\giga\eV}$~\cite{Arnaud:2017bjh}. The data from {\sc SEDINE} was also used to perform a search for Kaluza-Klein axions produced in the Sun, and a $90\%$ confidence level upper limit of $g_{a\gamma\gamma} < \num{8.99e-13}\,\si{\per\giga\eV}$
was set on the axion-photon coupling~\cite{NEWS-G:2021vfh}. 
{\sc S140}, a $\varnothing140\,\si{\centi\meter}$ spherical proportional counter made of 99.99\% pure copper is currently operating in SNOLAB, Canada~\cite{NEWS-G:2022kon}. First preliminary results provide the strongest constraint on spin-dependent WIMP-proton cross-section in the $0.2$-$2\,\si{\giga\eV}$ DM mass range.
{\sc S140}'s active volume is internally shielded with a $500\,\si{\micro\meter}$ thick layer of ultra-radiopure copper that has been deposited on the inner surface by adapting a low-background electroforming method to hemispheres. This procedure was undertaken at LSM~\cite{Knights:2019tmx,Balogh:2020nmo}.
Despite using $99.99\%$ pure copper and the electroformed internal shield, the dominant remaining background in {\sc S140} is the radioactive contamination and cosmogenic activation of the copper. Nevertheless, this can be mitigated by fully electroforming future detectors directly in the underground laboratory where they will be operated. 
This is the objective of Electroformed Cuprum Manufacturing Experiment ({\sc ECuME}), a $\varnothing 140\,\si{\centi\meter}$ spherical proportional counter that will be fully electroformed underground in SNOLAB. By fully electroforming the intact detector, additional radioactive contamination brought by machining and welding processes is avoided, and by conducting this underground, cosmogenic activation is minimised. 
 {\sc ECuME} will be operated with a neon--methane gas mixture (Ne:CH$_4$, 90\%:10\%) at 2~bar. 
The electroformed {\sc ECuME} detector will be installed in the shielding currently used for {\sc S140} upon conclusion of its physics exploitation.

To further reduce background rates, the detector shielding needs to be redesigned. In {\sc S140} a compact lead-based shielding is implemented, which provides the next largest background source after the detector construction materials~\cite{NEWS-G:2022kon}.
Additionally, recent advancements in spherical proportional counter read-out instrumentation are enabling the operation of larger and higher-pressure detectors, allowing the exposure to be greatly increased. 
These developments are the motivation behind \proposalAcronym, a proposed $\varnothing300\,\si{\centi\meter}$ spherical proportional counter electroformed underground at the Boulby Underground Laboratory.
\proposalAcronym\ will initially operate with a helium--isobutane gas mixture (He:i-C$_4$H$_{10}$, 90\%:10\%) at 5~bar and has the aim of reaching the neutrino floor in the DM-nucleon scattering cross-section for light DM.  The use of a gas which features a natural abundance of carbon-13, also provides sensitivity to the full range of DM effective field theory interactions~\cite{Fitzpatrick:2012ix}.

The \proposalAcronym\ design presented in this article is principally optimised to search for the light DM-nucleon interaction. However, the characteristics of the \proposalAcronym\ detector provide for a multi-physics platform: 
  the spherical proportional counter's single-electron threshold enables excellent sensitivity to the DM-electron interaction in the 10~MeV to~GeV mass range~\cite{Hamaide:2021hlp}; 
  and the large diameter of the spherical proportional counter enables sensitivity to ultra-heavy DM with masses close to the Planck mass~\cite{Bramante:2018qbc}.
Furthermore, in addition to DM searches, the energy resolution and light-readout capabilities of
a large spherical proportional counter filled with $^{136}$Xe gas lends itself to a neutrinoless double $\beta$-decay search, which is being explored by the Rare Decays with Radial Detector (R2D2) R\&D effort~\cite{Bouet:2020lbp,Katsioulas:2021usd}, 
and as a potential 
tool for supernova neutrino searches~\cite{Vergados:2005ny, Meregaglia:2017nhx}. 
 
This article, which presents the feasibility and the physics potential of this large volume, fully electroformed underground spherical proportional counter, is structured as follows. In Section~\ref{sec:DarkSPHERE}, the detector construction and operation, key expected performance characteristics and open R\&D topics are discussed. 
In Section~\ref{sec:Boulby}, the Boulby Underground Laboratory is presented as a proposed host of the \proposalAcronym\ experiment.
Section~\ref{sec:shielding} presents the conceptual design of the detector shielding and discusses the dominant background contributions. In Section~\ref{sec:Physics}, the physics potential of \proposalAcronym\ is discussed. We summarise our results and give our conclusions in Section~\ref{sec:summary}. 

\section{The DarkSPHERE Detector \label{sec:DarkSPHERE}}

This section discusses the key developments that will enable the discovery potential of the \proposalAcronym\ detector. We discuss the use of underground electroformed copper, the development of the multi-anode read-out,
quenching factor measurements, and the detector calibration methods.
An integral part of understanding the detector and characterising its capabilities is the state-of-the-art simulation framework for spherical proportional counters developed at the University of Birmingham~\cite{Katsioulas:2020ycw}.
The simulation framework combines several common physics simulation packages: Geant4~\cite{GEANT4} to simulate particle passage and interaction in matter; Garfield++~\cite{Veenhof:1998tt} for the simulation of the gaseous detector operation, interfacing with HEED for particle interactions; Magboltz~\cite{magboltz} for modelling electron transport parameters in gases; and ANSYS~\cite{ansys} for finite element method calculations of the electric field in the detector.  This simulation framework has been used to study detector calibration, fiducialisation and R\&D by the NEWS-G and R2D2 experiments~\cite{Giomataris:2020rna, Bouet:2020lbp}.

\subsection{Copper electroformation}

Copper is a common choice for a high-purity low-background material~\cite{BUCCI2004132,ARMENGAUD2010294,Aalseth_2018}
because of its commercial availability and the lack of long-lived radioisotopes -- $^{67}$Cu is the longest-lived with a half-life of $61.8\;\si{hours}$~\cite{nndcChartNuclides}. 
Even without long-lived radioisotopes, a sample of copper will contain some non-copper radiogenic contamination.
As an example, cosmogenic activation of the copper by cosmic-ray neutrons interacting through the
($n$,\,$\alpha$) reaction can produce $^{60}$Co, which, with a half-life of approximately 5.3$\;\si{years}$, is a long-lived background
relative to the typical timescale of rare event search experiments.
The copper industrial production processes also introduce radio-contaminants,
 primarily originating from the $^{238}$U and $^{232}$Th decay chains. It has been demonstrated that $^{222}$Rn introduced into the copper during manufacturing, and its progeny, are the dominant contaminants in terms of activity~\cite{ABE2018157}.

A method to produce ultra-radiopure copper is potentiostatic electroforming~\cite{ABE2018157,Hoppe2008}. This method takes advantage of electrochemical properties to produce copper with substantially reduced impurities. Electroformed copper has been produced with contaminant activity levels for $^{238}$U, $^{232}$Th and $^{210}$Pb below the current world-leading radio-assay techniques for these isotopes: inductively Coupled Plasma Mass Spectroscopy (ICP-MS) for $^{238}$U and $^{232}$Th, and the XIA UltraLo-1800 alpha spectroscopy for $^{210}$Pb~\cite{Abgrall:2016cct, ABE2018157,Bunker:2020sxw}.
Electroforming has been used to produce a variety of components for different rare-event search experiments, including the Majorana Demonstrator~\cite{Abgrall:2013rze} and NEWS-G~\cite{Balogh:2020nmo}. The electroplating of copper to {\sc S140}'s inner surface, performed in LSM, was the largest single-piece deep-underground electroformation ever conducted. 

Building on the experience gained with {\sc S140}, NEWS-G pursues project {\sc ECuME}: a deep underground electroforming facility, initially dedicated to electroforming a $\varnothing 140\,\si{\centi\meter}$ spherical proportional counter of the same name, but later could be used by other experiments. Given the scale of the {\sc ECuME} sphere, which will be the largest deep underground electroformed vessel, significant R\&D is ongoing to demonstrate the scalability of the spherical electroforming technique. Currently, work is ongoing to construct the scaled electroforming bath, which is similar in size to those used by the Majorana Demonstrator designed by Pacific Northwest National Laboratory (PNNL)~\cite{Abgrall:2013rze}, and includes detailed fluid dynamics and electrostatic calculations. The electroforming of a $\varnothing 30\,\si{\centi\meter}$ intact sphere at PNNL is ongoing, demonstrating the principle of intact sphere electroforming, and using only methods that are applicable to the larger scale of {\sc ECuME}.

The simplicity of the detector, comprising a single sphere, facilitates scaling of the electroforming, and this is an active area of research. 
The mandrel used to electroform the sphere is being designed, including the specific material used, the method of removing the mandrel from the interior of the electroformed copper sphere, and the structures required to support it. Additionally, the grounded rod supporting the read-out sensor will be electroformed, following established methods. All of this R\&D is directly applicable for \proposalAcronym\, and also includes auxiliary systems such as power-supply requirements and tolerances, assay and quality assurance techniques, and procedures for handling the electrolyte.

\begin{figure}[t!] 
  \centering
  \includegraphics[width=0.92\linewidth]{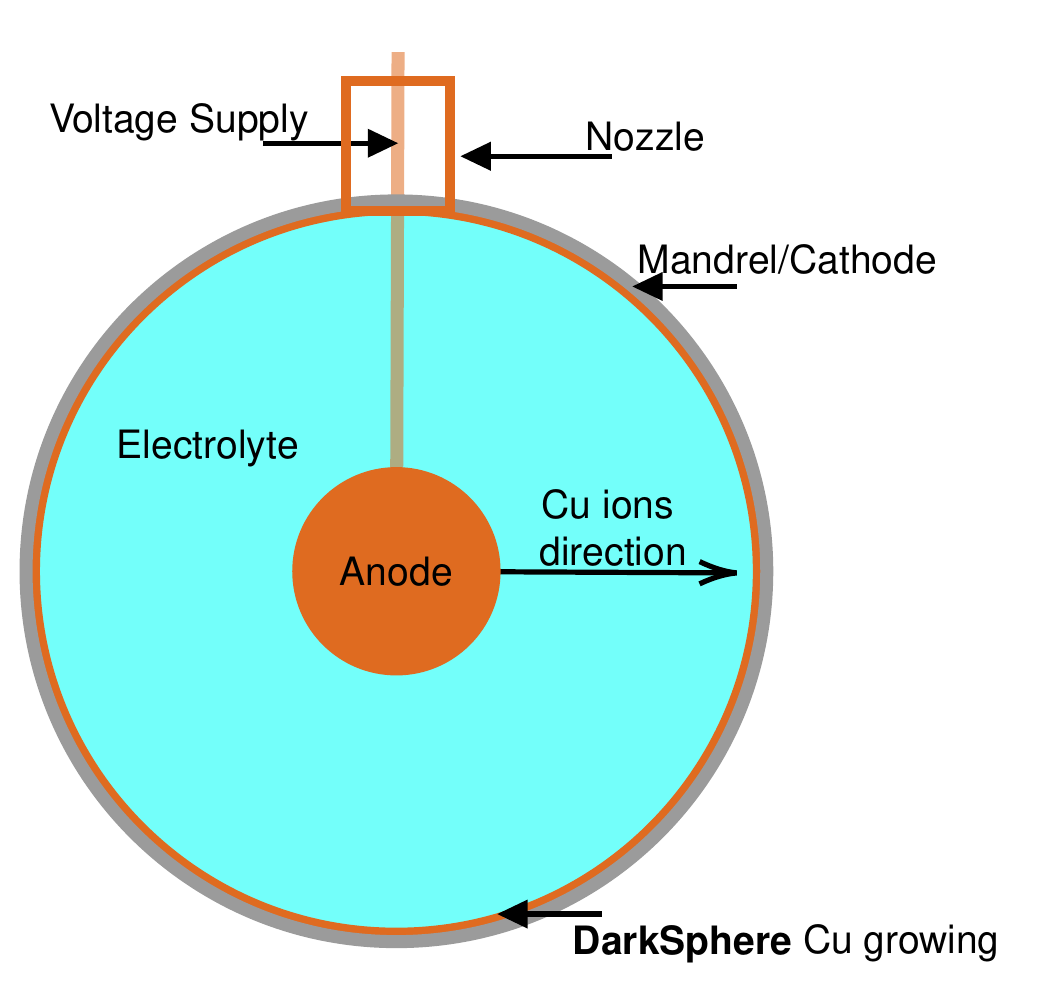}
  \caption{Conceptual schematic of the electroforming of \proposalAcronym. Due to its size, the electroforming will be performed in-to-out, with the cathode acting as the mandrel to which copper is deposited. The mandrel can then be chemically removed after the plating. The nozzle, which provides access to install the sensor during detector operation, will be electroformed directly in place. All surfaces which are in contact with the plating copper or the nozzle will be either submerged in electrolyte or in a nitrogen atmosphere.\label{fig:DarkSPHEREelectroforming}}
\end{figure}

Given the increased scale of \proposalAcronym, the electroforming may need to be conducted by plating to the inner surface of a mandrel, as shown in Figure~\ref{fig:DarkSPHEREelectroforming}. This has several advantages for both technical and radiopurity considerations. Firstly, this would allow the volume which the electrolyte and the produced copper surface are held to be hermetically sealed to the environment. While nitrogen cover gases are used in current electroforming projects to mitigate radon daughter deposition on the surface, this method would facilitate the use of a nitrogen atmosphere, suppressing radon daughter contamination. Secondly, in this in-to-out configuration, the mandrel would also act as the bath for the electrolyte, reducing the amount of electrolyte solution required. Methods for achieving this kind of electroforming are under investigation, but are similar to the basis of the NEWS-G electroforming~\cite{Balogh:2020nmo}, which used the hemispherical mandrel as the vessel.

  \begin{figure}[t!] 
  \centering
  \includegraphics[width=0.92\linewidth]{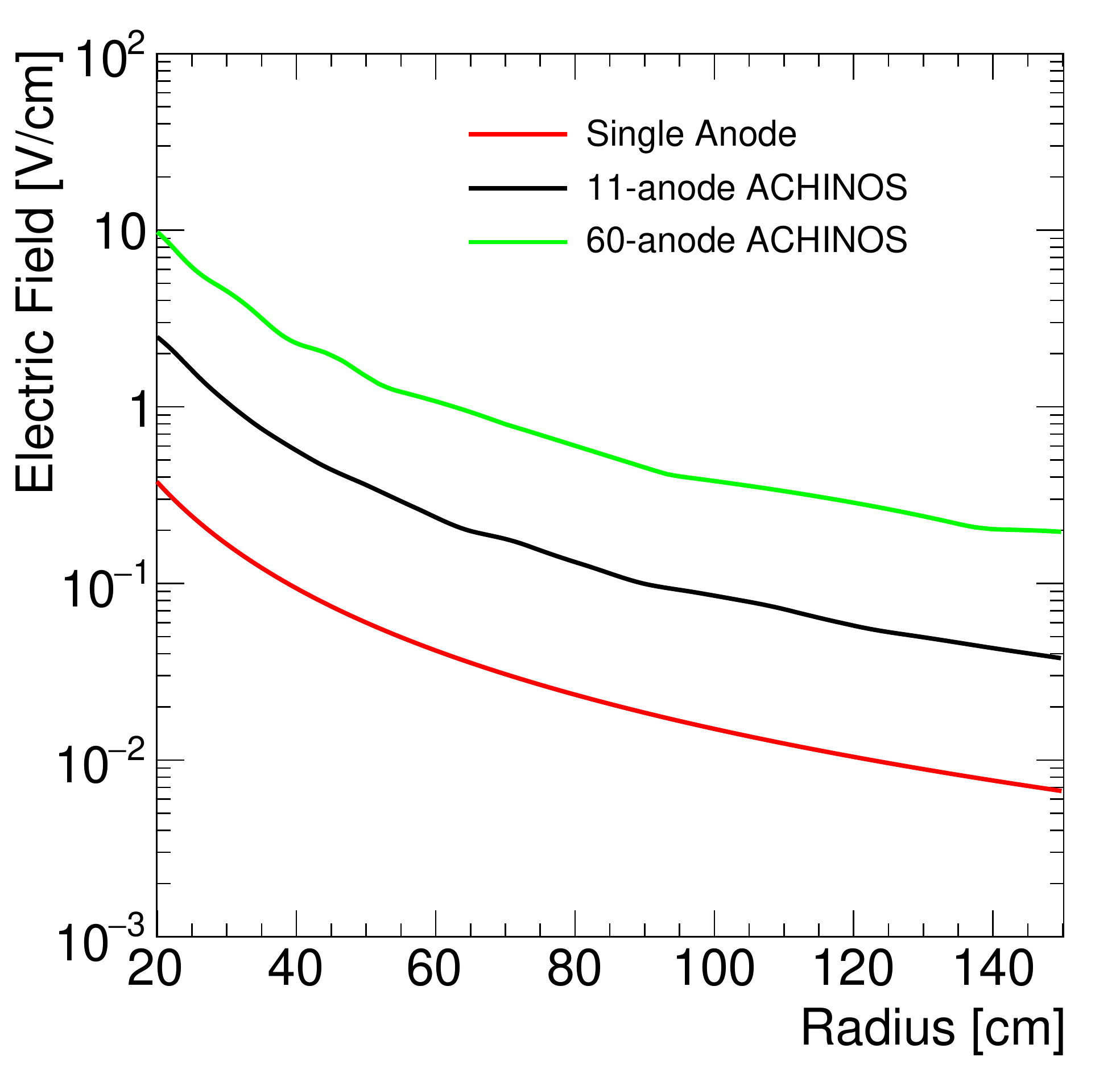}
  \caption{Electric field magnitude in the drift-region as a function of radius for different read-out configurations in \proposalAcronym. Each ACHINOS read-out uses $\varnothing 1\;\si{\milli\meter}$ anodes at the same voltage, and  $r_s=10\;\si{\centi\meter}$. Increasing the number of anodes increases the drift-region electric field magnitude without increasing the anode voltage.\label{fig:DarkSPHERE_sensorField}}
\end{figure}
\subsection{Sensor development}
\label{sec:sensorDev}

 The state-of-the-art read-out currently employed by NEWS-G, and spherical proportional counters in general, is the multi-anode `ACHINOS' structure. This is a scalable multi-anode sensor that allows for stable operation of larger detectors operating under higher pressures~\cite{Giganon:2017isb,Katsioulas:2018pyh, Giomataris:2020rna}.  The ACHINOS sensor, an example of which is shown in Figure~\ref{fig:schematicSPC}, comprises several spherical anodes located at positions equidistant from the detector centre. The sensor has been developed to decouple the electric fields in the drift and avalanche regions. This was not possible for single-anode sensors~\cite{Katsioulas:2018pyh}, which are suitable for smaller detectors or low-pressure operation. With ACHINOS, the drift region electric field depends on the collective field of all anodes while in the avalanche region it is determined by the field of the individual anode to which the ionisation electron arrives.

  \begin{figure}[t!]
     \centering
   \includegraphics[width=0.80\columnwidth]{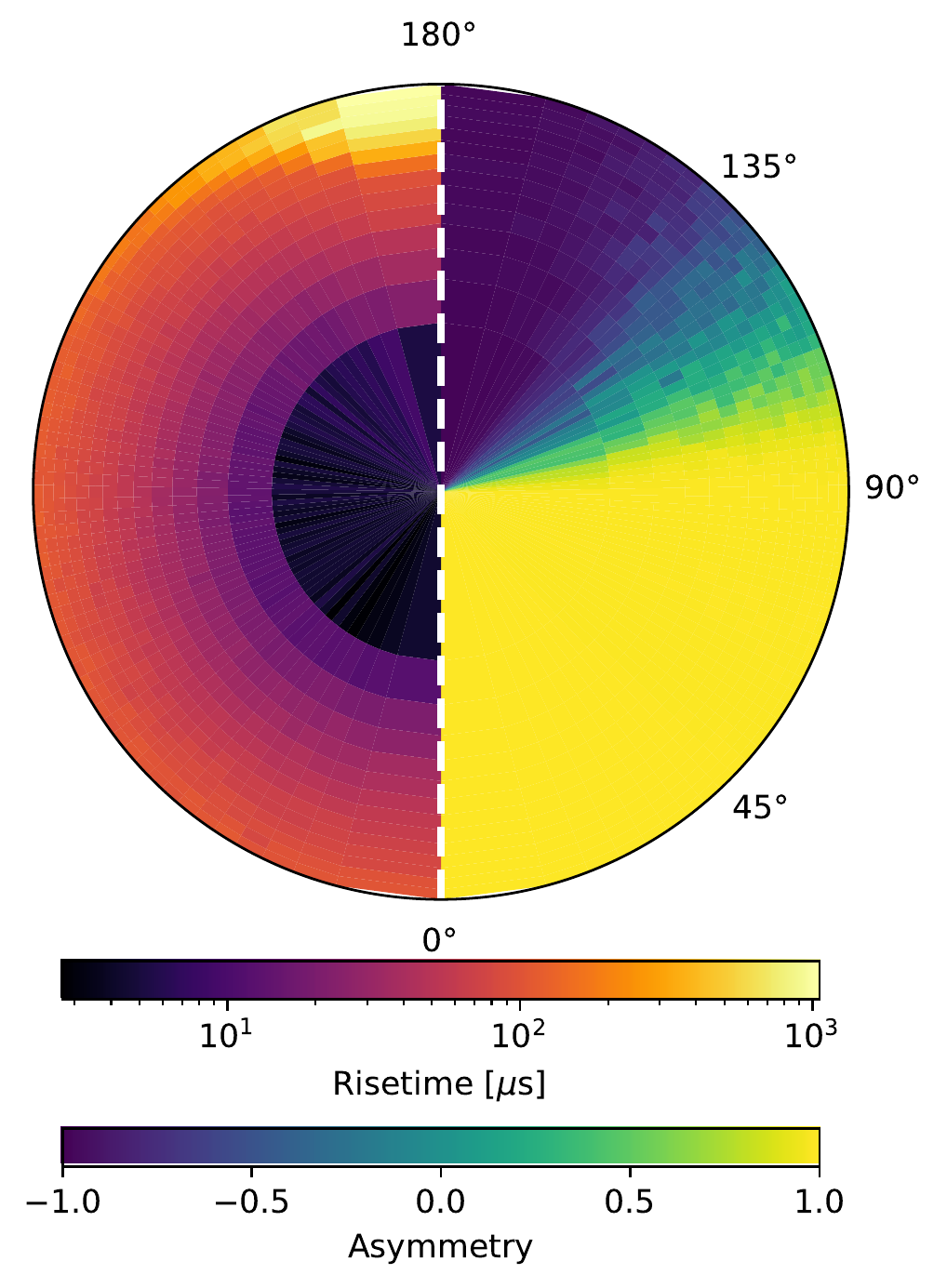}\\
     \caption{Simulation of $^{37}$Ar decays in a spherical proportional counter equipped with an 11-anode ACHINOS with two-channel read-out (Near and Far). Left side: the pulse rise time, which shows a correlation with the interaction radius. Right side: the Near-Far signal asymmetry, which 
     is correlated with the hemisphere of the detector in which the interaction occurred. Together, the pulse rise time and signal asymmetry provide information on the location of the interaction within the detector.
      \label{fig:achinosAsymmetry}}
\end{figure}

Current ACHINOS technologies use 11-anodes located at the vertices of an icosahedron, with the twelfth vertex being occupied by the grounded support rod, as shown in Figure~\ref{fig:schematicSPC}. Figure~\ref{fig:DarkSPHERE_sensorField} shows the increase by a factor of $6$ in the magnitude of the drift-region electric field that is achieved with an 11-anode ACHINOS compared to a single anode read-out at the same voltage.

In Figure~\ref{fig:achinosAsymmetry} the reconstruction of the interaction location is shown for the current 11-anode ACHINOS technology, which has a two-channel read-out grouped by the five anodes near the rod (Near) and the six further away (Far). The left semi-circle in Figure~\ref{fig:achinosAsymmetry} shows the simulated pulse rise-time (10\% to 90\% of pulse rising edge) as a function of interaction position for $^{37}$Ar decays, demonstrating sensitivity to the radial position. The right semi-circle in Figure~\ref{fig:achinosAsymmetry} shows the Near versus Far signal asymmetry, $(\text{Far}-\text{Near})/(\text{Far}+\text{Near})$, for $^{37}$Ar decays as a function of interaction location, which demonstrates the ability to determine the hemisphere of the interaction. Together, the risetime and signal asymmetry provides information on where the decay occurred within the detector.

\begin{figure}[t!] 
  \centering
  \includegraphics[width=0.70\linewidth]{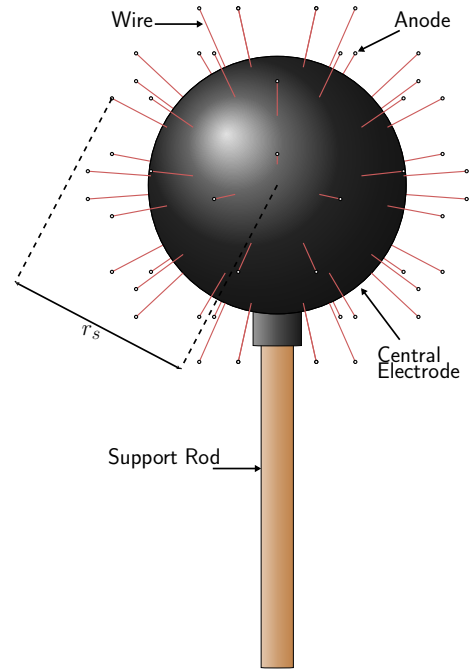}
  \caption{Schematic of an 60-anode ACHINOS where the anodes are located at the vertices of a truncated icosahedron. The distance between each anode and the centre is labelled as $r_s$. The multi-anode ACHINOS enables the operation of larger and higher pressure spherical proportional counters by decoupling the drift-region electric field from the avalanche electric field. \label{fig:60anodeACHINOS}}
\end{figure}

 \begin{figure}[t!] 
  \centering
  \includegraphics[width=0.60\linewidth]{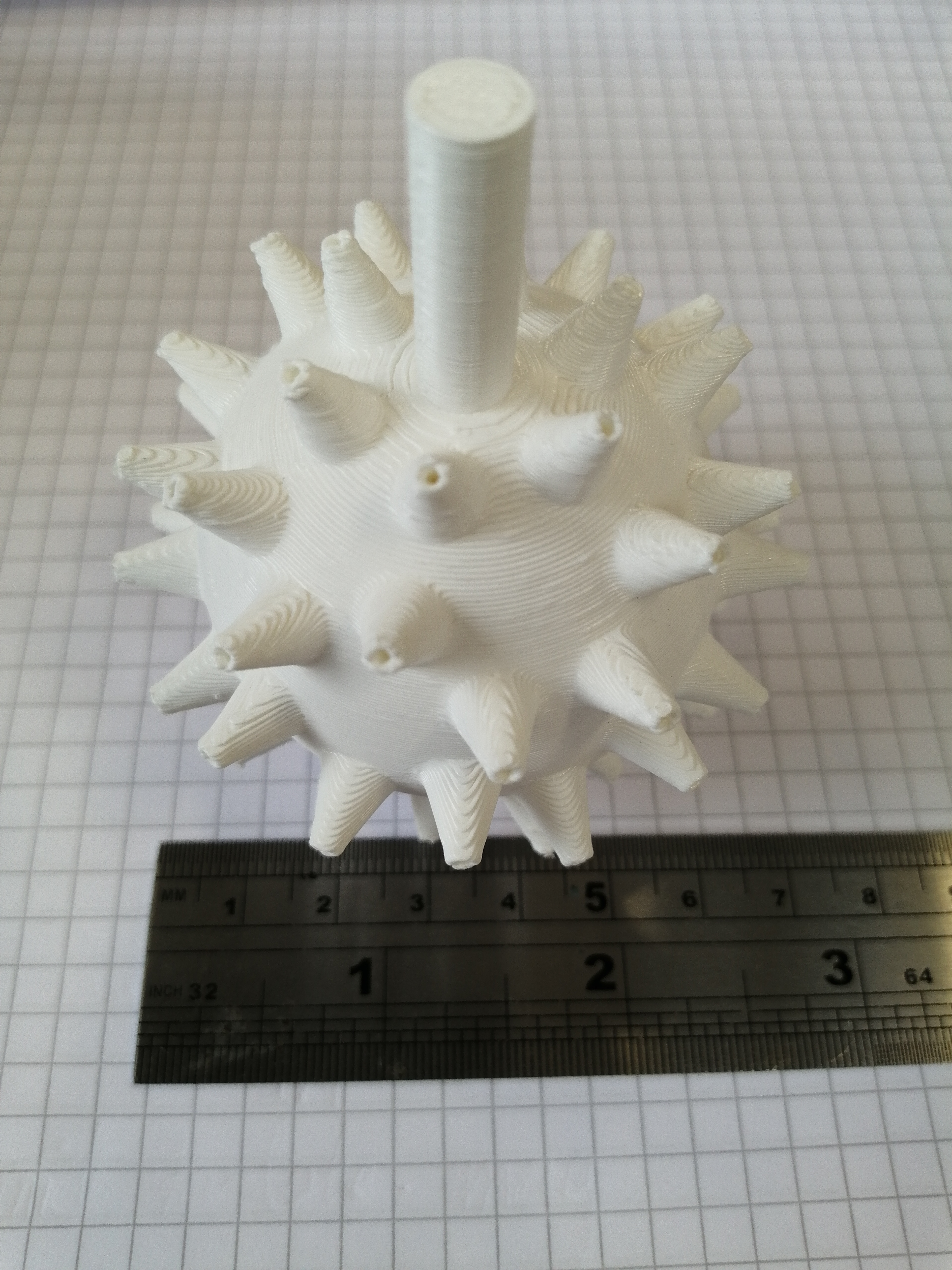}
  \caption{3D-printed, $25\%$ scale model of the prototype central electrode structure for the 60-anode ACHINOS for \proposalAcronym.\label{fig:60anodeACHINOS3d}}
\end{figure}

While the 11-anode structure is sufficient for {\sc S140} and {\sc ECuME}, the larger diameter and higher pressures envisaged for \proposalAcronym\ mean that an ACHINOS with a greater number of anodes is required. For this reason, a 60-anode ACHINOS, in which the anodes are located at the vertices of a truncated icosahedron, is planned for \proposalAcronym.  Figure~\ref{fig:60anodeACHINOS} shows a conceptual model of a 60-anode ACHINOS, which increases the electric field magnitude by a factor of $4$ relative to the 11-anode ACHINOS, as shown in Figure~\ref{fig:DarkSPHERE_sensorField}.
The result is that the field at the edge of the \proposalAcronym\ detector (at a radius of $150\,\si{\centi\meter}$) is similar to the field that the 11-anode achieves at the edge of the {\sc S140} and {\sc ECuME} detectors (at a radius of $70\,\si{\centi\meter}$), and is sufficient to maintain the anode voltages and avalanche fields that allow for stable operation of \proposalAcronym.

The design and construction of the 60-anode ACHINOS is an active area of research, with finite element calculations being performed to study the electric field configuration for different designs. R\&D for its construction is also underway, with a prototype 3D-printed, central support structure shown in Figure~\ref{fig:60anodeACHINOS3d}. In order to facilitate the increased cabling required to read and bias all 60-anodes compared to the current 11-anode ACHINOS, the diameter of the support rod will be increased from $4\;\si{\milli\meter}$ ($6\;\si{\milli\meter}$) internal (external) diameter to $8\;\si{\milli\meter}$ ($10\;\si{\milli\meter}$). This will also better support the larger mass of the sensor. 

 \begin{figure}[t!] 
  \centering
  \includegraphics[width=0.99\columnwidth]{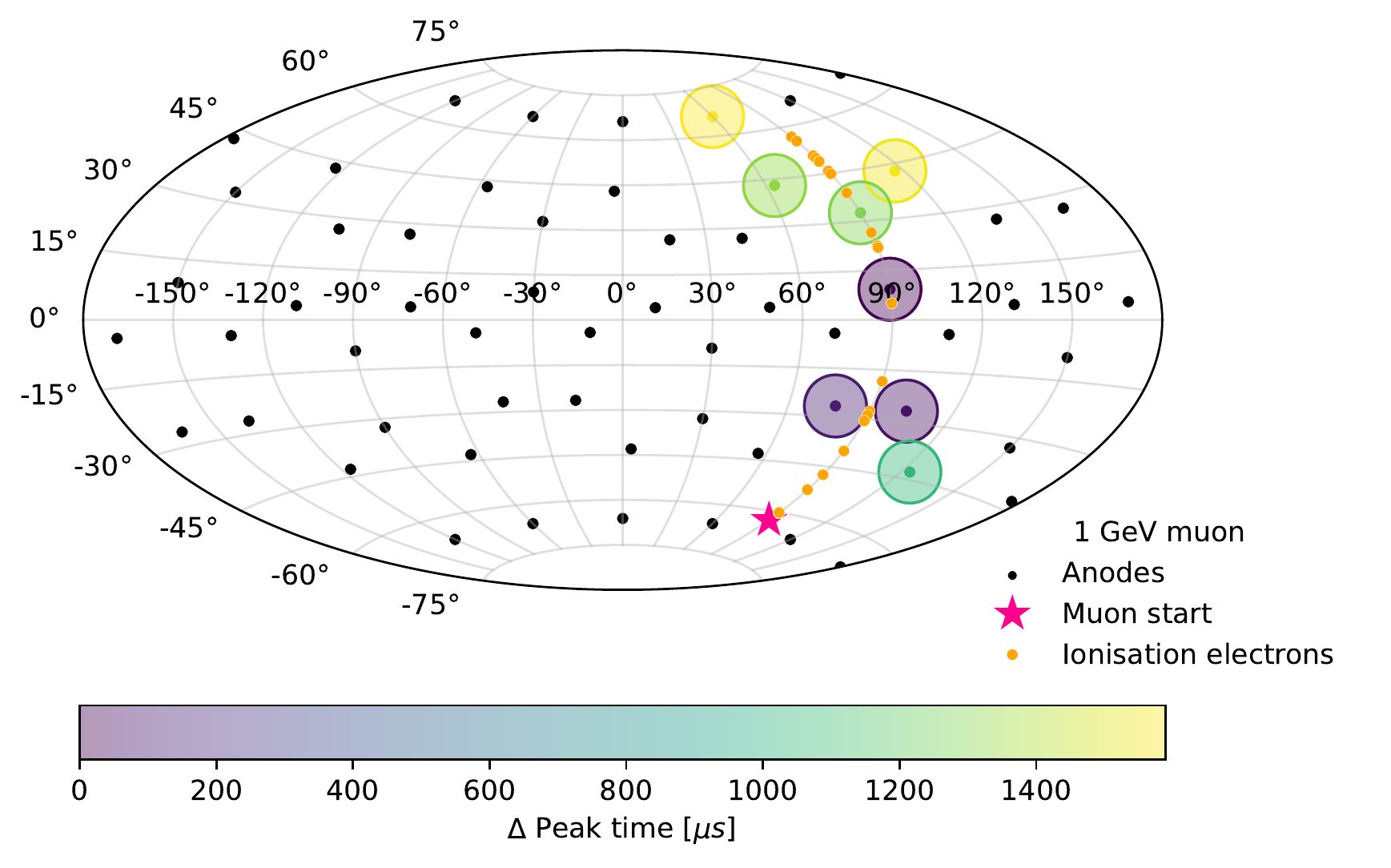}

  \includegraphics[width=0.8\columnwidth]{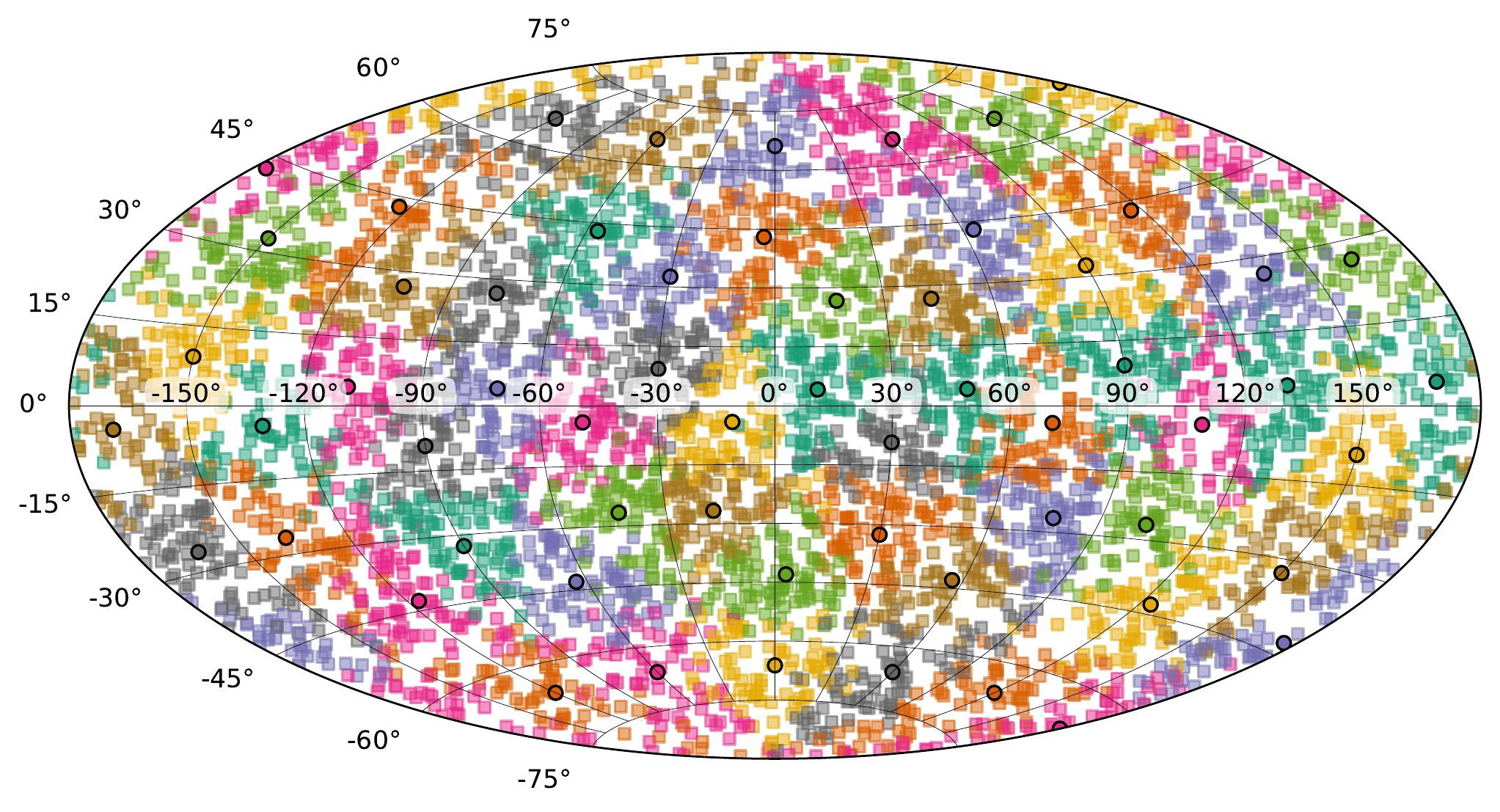}

  \caption{Simulated results for \proposalAcronym\ filled with isobutane and equipped with a 60-anode ACHINOS. The top panel shows a $1\;\si{\giga\eV}$ muon passing through the detector. The green star shows the starting position of the muon, the black points show the location of the 60 anodes, the orange circles show the position of ionisation electrons that are generated by the passage of the muon, and the larger coloured circles shows the relative peak time for the signal generated at each anode. There is a correlation between the muon's path and the peak time of the signal along the path. Pulse-shape properties are being explored to perform track reconstruction. The lower panel shows a simulations of single electrons uniformly distributed inside \proposalAcronym. The crosses show the initial electron position, and the colour indicates to which anode that electron arrived, with the anodes shown as colour dots. \label{fig:achinosMuon}}
\end{figure}

The individual anode read-out of the ACHINOS sensor enhances the potential for position-sensitive interaction information. This gives \proposalAcronym\ the potential to significantly improve upon current 2-channel 11-anode ACHINOS read-out, by using a 60-anode ACHINOS with individual anode read-out. This is demonstrated in Figure~\ref{fig:achinosMuon}, where the upper panel shows a simulated $1\;\si{\giga\eV}$ muon passing through \proposalAcronym, while the lower panel shows the reconstructed positions of single electrons. In both cases, the simulations assume a detector filled with isobutane and equipped with a 60-anode ACHINOS. The peak time of the signal in each anode is used to infer the track direction, and work is ongoing to utilise further pulse properties to perform track reconstruction.
This development offers improved detector fiducialisation and background rejection capability, and may enable the search for DM candidates that leave tracks in the detector (discussed further in Section~\ref{sec:Physics}).

\subsection{Quenching factor measurements \label{sec:quenching}}
A nuclear recoil induced by the elastic scattering of DM with a target nucleus in the gas will dissipate only a fraction of its energy as ionisation, known as the quenching factor. This is the fraction of the recoil energy that will be observable by the spherical proportional counter. Knowledge of the quenching factor is essential to reconstruct the recoil energy from the observed ionisation signal.  

The NEWS-G Collaboration is actively pursuing several methods of measuring the quenching factor for low-energy ions in gases. Firstly, the COMIMAC facility at LPSC Grenoble~\cite{Muraz:2016upt} uses a compact Electron Cyclotron Resonance (ECR) source to produce either ions or electrons of various energies, selected by a tunable extraction potential. The electrons or ions are directed into a gaseous detector volume, which can be either a spherical proportional counter or a MicroMegas~\cite{Giomataris:1995fq}, where they induce ionisation. The quenching factor can be extracted by comparing the measured signal from the electrons to the signal when ions are used. This method has previously been used to measure the quenching factor of He$^{+}$ in He/C$_{3}$H$_{8}$ gas mixtures, and protons in i-C$_4$H$_{10}$/CHF$_3$~\cite{Tampon:2017mpm,santos2008ionization} and CH$_4$~\cite{NEWS-G:2022fym}.  Measurement campaigns are ongoing or envisaged for other gases including those proposed for use in \proposalAcronym.

A second method is to induce nuclear recoils using the scattering of neutrons of known energy, as employed at TUNL, USA~\cite{NEWS-G:2021mhf}. Pulsed bunches of $20\;\si{\mega\eV}$ protons produced by a tandem Van de Graaf accelerator are directed onto a lithium fluoride target, undergo the $^{7}$Li(p, n)$^{7}$Be reaction, and produce monochromatic neutrons. The neutrons then scatter in the gaseous target, which is within a spherical proportional counter, and go on to interact in a scintillator backing detector. Different recoil energies were selected by changing the angle between the lithium fluoride, the spherical proportional counter and the backing detector. The quenching factor is reconstructed by comparing the signal generated by the recoiling nucleus to the detector calibration performed using radioactive sources, such as $^{55}$Fe. Measurements demonstrating the principle have been concluded~\cite{NEWS-G:2021mhf}, and further measurements are ongoing with the gases envisaged for \proposalAcronym. 

A third method using existing measurements of the $W$-value of electrons and ions in gases has also been developed~\cite{Katsioulas:2021sgl}. $W$-values have been extensively studied in the context of dosimetry so have focused on tissue-equivalent gases, their constituents, and other common gases. They are generally measured using ionisation chambers where care is taken to mitigate any detector-specific effects that would reduce the generality of the measured $W$-values. While the energy loss for electrons is dominated by electronic effects, which results in ionisation, the energy loss of ions has contributions from other processes such as excitation that do not produce ionisation. Therefore, the relative comparison of the two sets of $W$-value measurements provide an estimate of the quenching factor over a given energy range. This method can be used to provide quenching factors in additional gases as further $W$-value measurements become available.

\subsection{Detector calibration}

\begin{figure}[t!] 
  \centering
  \includegraphics[width=0.95\linewidth]{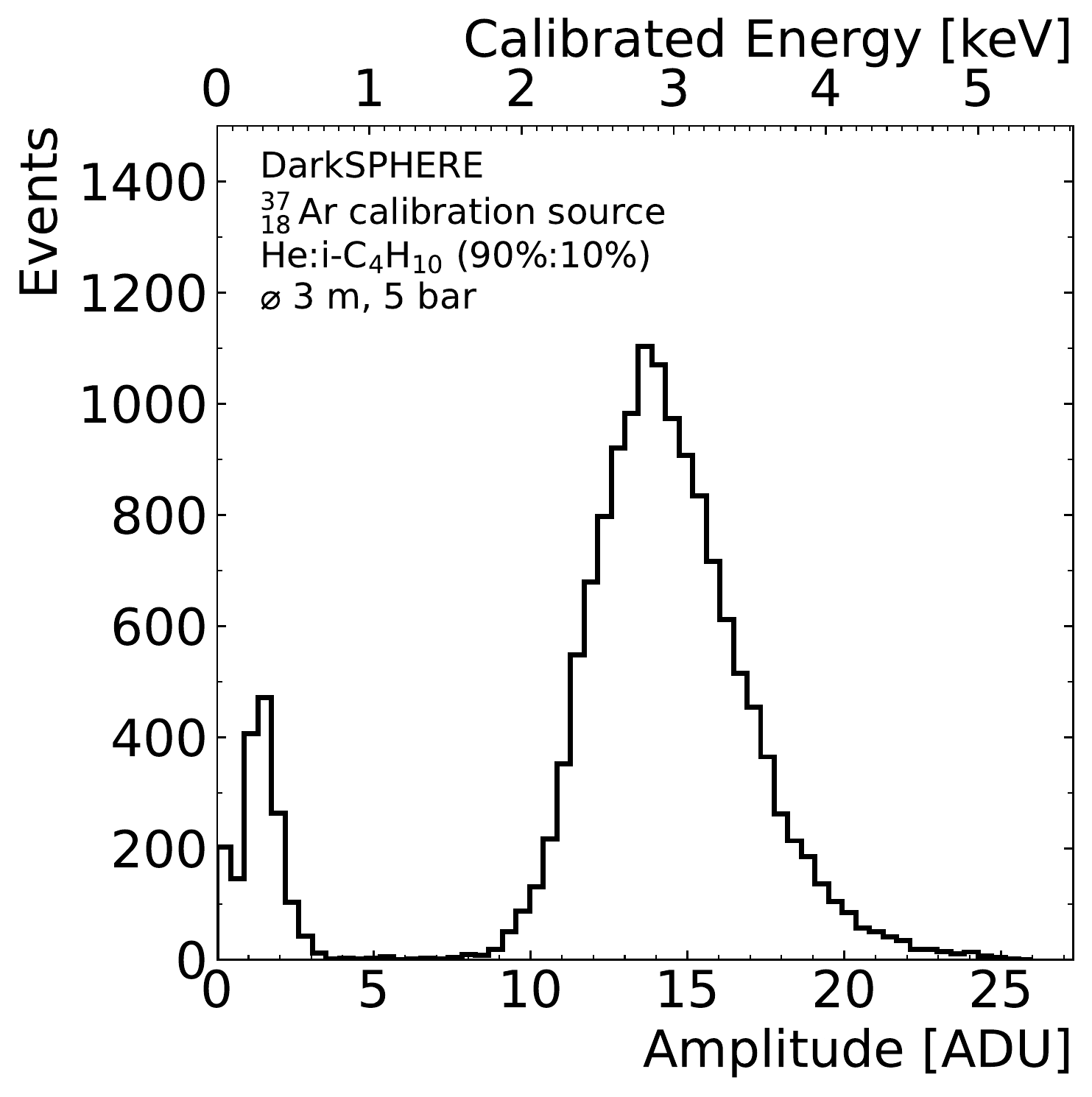}
  \caption{Simulated energy spectrum recorded by \proposalAcronym\ equipped with a 60-anode ACHINOS, with $r_s=30\;\si{\centi\meter}$, when an $^{37}$Ar gaseous calibration source is placed inside the detector. The peaks at 270~eV and 2822~eV correspond to the L- and K-shell decays of $^{37}$Ar.
   \label{fig:Ar37}}
\end{figure}

\proposalAcronym\ will be calibrated using methods developed by NEWS-G, which are outlined in Ref.~\cite{Arnaud:2019nyp}.
A combination of a UV-laser system and a
gaseous $^{37}$Ar source allows the calibration and characterisation of the detector. 
Laser data is used to extract electron avalanche parameters, namely the mean gain, $\langle G\rangle$, and the Polya distribution parameter, $\theta$, Fano factor $F$, and gas properties, such as electron drift times. The laser also provides continuous online monitoring of the detector operation during data taking. Data
from $^{37}$Ar decays will be used to measure the 
$W$-value of the target gas at 270~eV and 2822~eV,
corresponding to the L- and K-shell decays of
$^{37}$Ar, which will occur uniformly throughout the detector. Figure~\ref{fig:Ar37} shows an example of the energy spectrum that would be measured in \proposalAcronym\ equipped with a 60-anode ACHINOS when an $^{37}$Ar source is inside the detector. The two energy lines
are clearly visible. To provide more energies for the calibration, other radioisotopes or an X-ray generator~\cite{Giomataris:2020rhd} could be used in conjunction with a window in the detector. Additionally, the detector will be calibrated for nuclear recoils using a neutron source, for example $^{241}$Am-$^{9}$Be.

\subsection{In-situ neutron measurements}
\label{sec:neutrons}
Neutrons originating from radioactive decays or induced by cosmic-ray muons are a part of the background in underground facilities and should therefore be characterised and mitigated. For \proposalAcronym, the neutron induced background will be measured in-situ by using a nitrogen-based gas mixture, exploiting the reactions
\begin{eqnarray}
\nonumber ^{14}\mathrm{N} + n \rightarrow\,\! ^{14}\mathrm{C} + p + 625\;\si{\kilo\eV}\,, \\
\nonumber ^{14}\mathrm{N} + n \rightarrow\,\! ^{11}\mathrm{B} + \alpha - 159\;\si{\kilo\eV}\,.
\end{eqnarray}
The $(n,p)$ interaction has a thermal cross-section of $1.83\,\si{\barn}$,
while the $(n,\alpha)$ reaction becomes significant for neutron energies above $2\;\si{\mega\eV}$.

A first demonstration has proven the feasibility of the method~\cite{Bougamont:2015jzx}, and is an active area of investigation~\cite{Giomataris:2021fwv,Giomataris:2022kxw}. Recently, using the sensor developments described in Section~\ref{sec:sensorDev},
an 11-anode ACHINOS sensor with $\varnothing1\,\si{\milli\meter}$ anodes and read-out with the Near- and Far-channels~\cite{Giomataris:2020rna}, was installed in a $\varnothing30\,\si{\centi\meter}$ spherical proportional counter at the University of Birmingham. 
An $^{241}$Am-$^{9}$Be neutron source 
provided fast neutrons, which could be thermalised using a graphite stack. Fast and thermal neutrons were detected and characterised with a spherical proportional counter operating with up to $1.8\;\si{\bar}$ N$_{2}$ gas~\cite{Giomataris:2022bvz}.

\section{Boulby Underground Laboratory}
\label{sec:Boulby}

The Boulby Underground Laboratory was established in 1987 and is located at a
depth of $1100\,\si{\meter}$, equivalent to $2840\,\si{\meter}$ of water.   
The laboratory is operated by
the UK’s Science and Technology Facilities Council (STFC) and has a track record in the development and support of low-background
physics, including the ZEPLIN dark matter programme which operated a series of three xenon detectors until
2011~\cite{Alner:2005pa,Alner:2007ja,Akimov:2006qw,Lebedenko:2008gb}, and the DRIFT and CYGNUS directional DM programmes~\cite{Daw:2010ud,Battat:2014van,Battat:2016xxe}.
DRIFT operated with a gas mixture of carbon disulfide, CS$_2$, and
carbon tetrafluoride, CF$_4$, providing the laboratory with unique
expertise in safe handling and operation of highly flammable and toxic
gases. Further experience with gaseous detectors and gas handling has been established through NEWS-G's 
collaboration with Boulby: a $\varnothing30\,\si{\centi\meter}$ spherical proportional counter is currently operated at Boulby
for spherical proportional counter instrumentation R\&D in a controlled environment~\cite{Katsioulas:2019iea}.

In 2015, a new laboratory area was constructed
with a $4000\,\si{\meter\cubed}$ experimental
space divided into a main hall and a connected area
known as the Large Experimental Cavern (LEC).
These areas are certified to ISO class~7 clean-room standard and are
serviced by cranes that facilitate material handling.  A further area 
is maintained at ISO class~6 standard and is dedicated
to the Boulby Underground Screening facility (BUGS)~\cite{Scovell:2017srl}. BUGS comprises six primary
ultra-low-background HPGe detectors and a small pre-screening detector for qualitative
measurements of materials suspected to be of high activity. Since
2015, BUGS has primarily supported the LUX-ZEPLIN (LZ) construction
material radio-assay campaign~\cite{Akerib:2015cja, Mount:2017qzi}, but has also performed assays for the SuperNEMO~\cite{Piquemal:2006cd} 
and
Super-Kamiokande~\cite{Mori:2013wua} experiments. 
Materials screening capabilities at Boulby have significantly expanded with the
installation of two UltraLo-1800 $\alpha$-spectrometer modules, which
can be used to estimate very low $^{210}$Pb contamination in copper bulk~\cite{Abe:2017jzw}, as well as with Radon Emanation detectors.%

Similarly to the {\sc ECuME} facility, STFC has awarded funding for a deep underground electroforming facility in Boulby. This will be a general-purpose electroforming facility capable of producing ultra-pure copper components for experiments in the laboratory and internationally. At the time of writing, procurement of equipment is underway, which will be followed by the commissioning of the smaller of two planned electroforming baths to be commissioned during 2023.  

The area surrounding the laboratory exhibits low seismic activity, and human-induced seismic activity is low, since mining generally does not use explosives and is more than $1\,\si{\kilo\meter}$ from the laboratory.
The geology of the cavern rock around the laboratory contributes to
its suitability for low-background activities.  The halite rock has
been measured to contain $(32\pm3)\;\si{ppb}$ of $^{238}$U, $(160\pm 20)\;\si{ppb}$ of $^{232}$Th, and (0.036$\pm$0.003)\% of {$^{40}$K}
\textcolor{purple}~\cite{Malczewski:2013lqy}.  The low level of $^{238}$U
contributes to a low ambient background from airborne $^{222}$Rn of
only $2.4\;\si{Bq\per\meter\cubed}$~\cite{Araujo:2011as}, which is significantly
lower than the values measured in other facilities~\cite{Ianni:2017vqi}.

\section{Shielding design and dominant backgrounds}
\label{sec:shielding}

\begin{table*}[t!]
\caption{Environmental background rates below $1\,\si{\kilo\eV}$ in the \proposalAcronym\ active volume with the $2.5\;\si{\meter}$ full-water shielding system. Environmental backgrounds are induced by photons, neutrons and cosmic-ray muons. High-energy photons that originate from decays in the cavern of the laboratory give the largest contribution to the total rate. The neutrons, which are produced directly from decays in the cavern and indirectly from interactions of cosmic-ray muons with the cavern rock, induce both neutrons and photons in the active volume and give the next largest rate. Contributions from muons interacting in the detector and its shield are also given, but these can be suppressed through active veto techniques.\label{tab:summaryResults} }
\begin{tabular}{c c c c c}
\toprule
\multicolumn{1}{c}{\textbf{}}          
&\multicolumn{4}{c}{\textbf{Environmental background rate $\bm{\leq 1}$\,keV [dru]}} \\ 
\multicolumn{1}{c}{\textbf{Shielding}}      
& \multicolumn{1}{c}{\textbf{  Photon-induced  }}  
&\multicolumn{2}{c}{\textbf{  Neutron-induced  }}  
&\multicolumn{1}{c}{\textbf{  Muon-induced}}\\ 
\textbf{Configuration}  & Photon & Neutron & Photon &   \\ 
\midrule
$2.5\,\si{\meter}$ water                            
& $\num{4.2e-3}\, (0.3)\;$ 
& $\num{9e-5} \,(5) $
& $\num{1.3e-4} \,(0.4)\; $ 
& $\num{5e-3} \,(4)\; $\\
\bottomrule
\end{tabular}
\end{table*}

\begin{figure}[t!]
  \centering
  \includegraphics[width=1\linewidth]{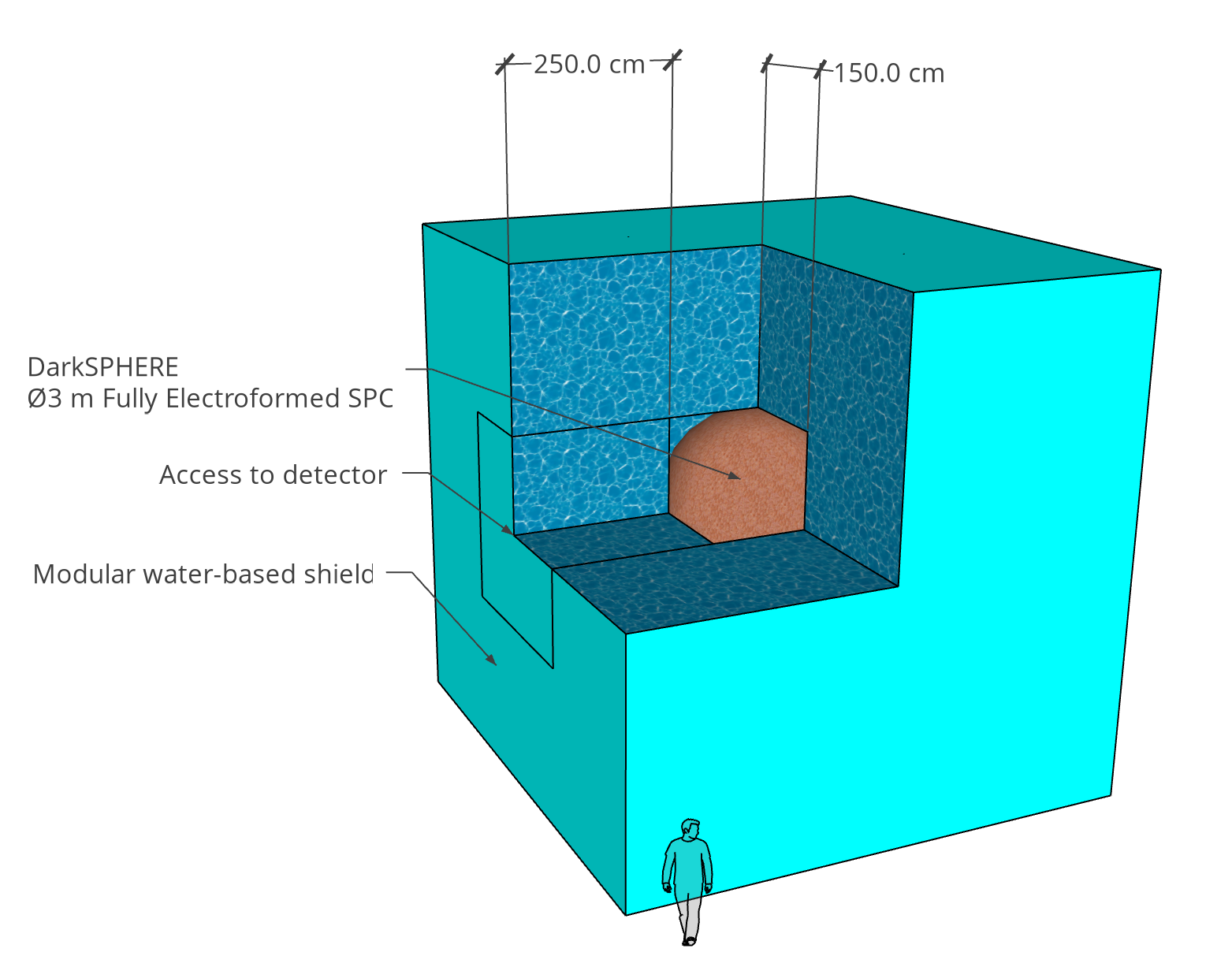}
  \caption{Schematic representation of the conceptual shielding design used to estimate background rates for \proposalAcronym. The $\varnothing3\,\si{\meter}$ electroformed-copper sphere is surrounded by a $2.5\,\si{\meter}$ water shield and will be housed in the Large Experimental Cavern (LEC) at the Boulby Underground Laboratory. 
   \label{fig:DarkSPHERE_shield} }
\end{figure}

The conceptional design for the \proposalAcronym\ shielding system assumes that the Boulby LEC will be the host location. The \proposalAcronym\ shielding should achieve a significant reduction in the background rate compared to the {\sc S140} and {\sc ECuME} experiments, which are expected to achieve background rates of $1.7\;\si{dru}$ and $0.3\;\si{dru}$, respectively~\cite{NEWS-G:2022kon}.
This aim can be achieved with a full-water shield, with a preliminary design shown in Figure~\ref{fig:DarkSPHERE_shield}. 
The full-water shield is a modular hollow-cube with a thickness of $2.5\;\si{\meter}$, and has the advantage that it is low-cost, straightforward to implement, and sufficiently pure water can be procured so that it does not provide an additional source of background. 
The water will be held in plastic containers, the specific material and design being selected based on structural and radiopurity considerations.
This design suppresses environmental backgrounds without introducing a significant background rate from radioactive contaminants in its materials. 

A Geant4 simulation~\cite{GEANT4} was used to study the experimental background originating from the laboratory environment, the detector and shielding materials, and the gas used. The simulation calculates the probability that a particle deposits an energy less than $1\;\si{\kilo\electronvolt}$ in the detector's active volume, and takes measured fluxes at Boulby of neutrons, photons, and muons as input.
To improve the efficiency of the simulation, a forced collision biasing scheme for Monte-Carlo variance reduction was used~\cite{Allison:2016lfl}. Biasing is applied to the three primary particles from the environmental background: photons, neutrons, and muons. For the background simulations, a thickness of $1\;\si{\centi\meter}$ was assumed for \proposalAcronym's electroformed cathode. Specific details of the simulation method for each considered background are discussed in the following sections. 

\subsection{Environmental Backgrounds}

The environmental background comprises neutrons and photons originating from decays in the cavern of the laboratory, neutrons produced by interactions of cosmic-ray muons with the cavern rock, and cosmic-ray muons directly interacting in the active volume and shielding. 

The background induced by neutrons and photons was simulated by assuming that the neutrons and photons originated with a random initial direction and a random position just outside an external face of the shield.
For the photons, the initial energy used in our simulations was drawn from the measured total photon flux as a function of energy from Ref.~\cite{Scovell:2017srl}.
For the neutrons, the initial energy was sampled from the measured energy distribution from Ref.~\cite{Smith:2005se},
which includes neutrons from radioactive decay and cosmic-ray muon interactions in the cavern rock (see Ref.~\cite{Kudryavtsev:2003aua}).

Table~\ref{tab:summaryResults} summarises the different environmental background components induced by neutrons, photons and muons. 
The dominant contribution to the environmental background rate from photons and neutrons below $1\;\si{\kilo\electronvolt}$ arises from the highest energy photons, from $2000-2750\;\si{\kilo\electronvolt}$. 
With a $2.5\;\si{\meter}$ water shield, the total background induced by neutrons and photons is $4.46\times10^{-3}$\,dru (events/kg/day/keV).

The muon-induced rate presented in Table~\ref{tab:summaryResults} is similar in magnitude to the total neutron and photon rate but it is envisaged that the muon-induced contribution can be further mitigated. This is because muons entering the detector volume will leave an extended track of ionisation electrons, as shown in the simulation in Figure~\ref{fig:achinosMuon}. 
Pulse-shape information and the multi-anode readout discussed in Section~\ref{sec:sensorDev} can then be used to discriminate against these events. 
In addition, muons interacting with the shielding or detector material may produce secondary particles that can interact in the detector. 
Several muon veto techniques are being explored to suppress this background, including for example, instrumenting the water-shield with light-sensitive readouts to detect Cherenkov radiation, as used in other experiments~\cite{DEAP-3600:2017ker}.

\subsection{Radioactivity from detector and shielding materials}

The dominant background contribution from the electroformed copper proposed for \proposalAcronym\ will be from $^{210}$Pb and progeny decays. Previous electroformed copper samples have demonstrated a contamination of $<0.12\;\si{\micro\becquerel\per\kilo\gram}$~\cite{ABGRALL201622}.
To assess the contribution of
$^{210}$Pb to the background rate, it was simulated uniformly in the $1\,\si{\centi\meter}$ thickness of the detector shell, with the decay chain modelled using the \texttt{Geant4RadioactiveDecayPhysics} package. The probability that a $^{210}$Pb decay causes an interaction depositing energy below $1\;\si{\kilo\eV}$ was found to be $\num{2.02e-5} \,(0.02)$. This results in a background event rate of $\num{1.90e-5}\,\si{dru}$, which is significantly smaller than the environmental background rates given in Table~\ref{tab:summaryResults}.

The simplicity of the shielding allows for a limited set of material to be used, which can be selected for their radiopurity. This includes the polyethylene used to hold the water shielding. Possible radioactive contaminants of the water were considered, for example, $^{40}$K, but were found to bring a negligible contribution to the overall background rate~\cite{ARNQUIST201715}. 

The background contribution of the read-out structure and it's support rod have not been considered at this time. As previously mentioned, the support rod will also be electoformed, and represents a negligible mass in comparison to the detector shell. The remaining components, being the central electrode, wires, and anodes of the ACHINOS, will be considered during the R\&D for the project, which will be based on current R\&D performed for NEWS-G. 

\subsection{Radioactivity from the gas mixture} 
The principal gas mixture to be used for the physics exploitation of \proposalAcronym\ is He:i-C$_{4}$H$_{10}$ ($90\%$:$10\%$). These gases are produced from underground natural gas deposits: He directly, making up a portion of natural gas; and i-C$_{4}$H$_{10}$ indirectly, being produced from the isomerisation and fractionation of butane, which is found in natural gas deposits. As such, these gases have been trapped underground for significant, geological time scales and so have no appreciable contamination by isotopes produced through capture reactions with cosmogenically-produced spallation neutrons, for example $^{14}$C or tritium. It is only during subsequent manipulation or purification where these isotopes can be produced. By working with gas manufacturing companies, it is possible to  ensure that gases that have spent the least time exposed to surface-level neutron fluxes can be used. This will minimise the abundance of cosmogenically produced radioisotopes present in the gas~\cite{Amare:2017roa}. Furthermore, some commercial gas purifiers, which are required to remove trace levels of oxygen and water from the gases, have been shown to emanate $^{222}$Rn. Methods to develop purifiers with greatly reduced $^{222}$Rn are being developed~\cite{altenmuller:2021fly} and will be employed for \proposalAcronym, along with current R\&D work ongoing in NEWS-G for radon traps.

\section{Physics Opportunities \label{sec:Physics}}
The large size and the low background of \proposalAcronym\ 
enables multiple searches for signals from phenomena beyond the Standard Model of particle physics. 
In this section, we begin by discussing the main searches for which spherical proportional counters are used,  namely, to search for DM with
a mass of $\mathcal{O}(\mathrm{GeV})$
that elastically scatters with a light target-nucleus. To demonstrate the multi-physics potential of \proposalAcronym\ we discuss how even heavier mass DM candidates that interact with an atomic nucleus could be constrained, before turning our attention to lighter mass DM candidates that interact with electrons. 
Finally, we consider the possibility of extending the use of spherical proportional counters by filling with $^{136}$Xe gas to search for neutrinoless double $\beta$-decay, and coherent neutrino-nucleus scattering.

\subsection{Nuclear recoils from DM scattering \label{sec:PhysicsNR}}

The baseline scenario is that \proposalAcronym\ will be operated with a He:i-C$_{4}$H$_{10}$ (90\%:10\%) gas mixture at a pressure of $5\,\si{\bar}$ for a total mass of $27.3\,\si{\kilo\gram}$. The use of He:i-C$_{4}$H$_{10}$ provides a significant amount of low mass nuclei, specifically hydrogen and helium, which means that DM with a mass around the GeV-scale can efficiently transfer kinetic energy to the nuclei.
Combined with the low-energy threshold of the spherical proportional counter gives \proposalAcronym\
 high sensitivity to low-mass DM candidates.

\begin{figure}[t!] 
  \includegraphics[width=0.95\columnwidth]{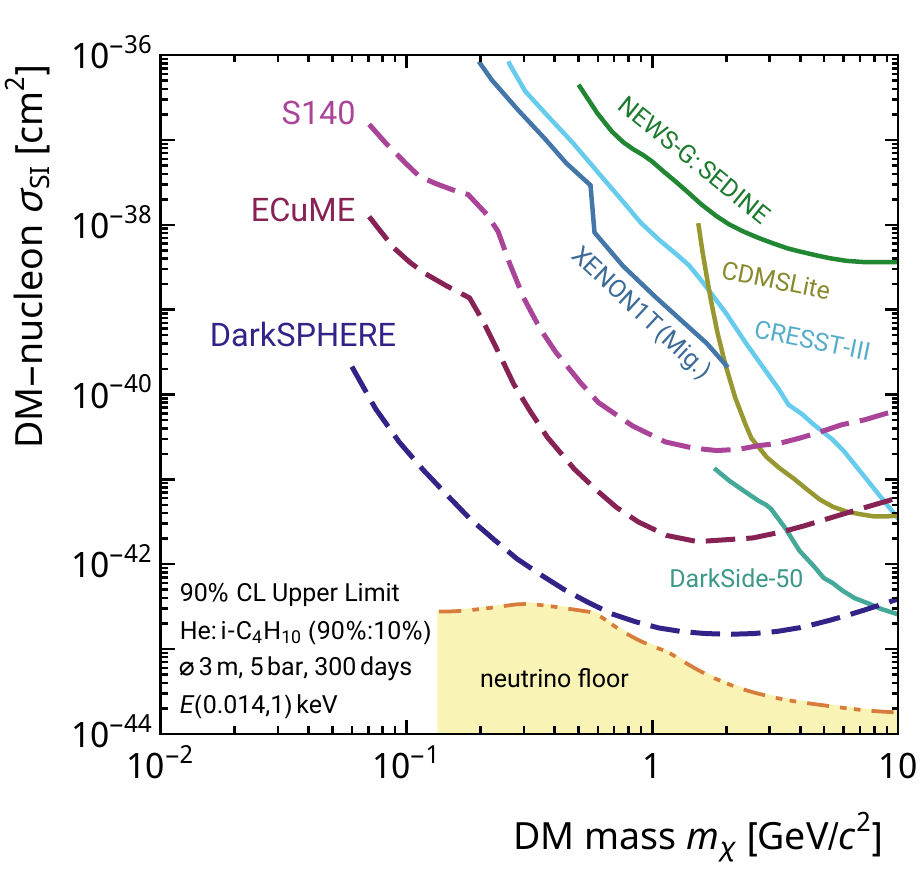}
  \caption{\label{fig:limits}
  Sensitivity projections (dashed) for {\sc S140}, {\sc ECuME} and \proposalAcronym\ to the spin-independent DM-nucleon cross-section.
  Solid lines show existing constraints from CDMSlite~\cite{Agnese:2018gze}, CRESST-III~\cite{Abdelhameed:2019hmk}, DarkSide-50~\cite{Agnes:2018ves},  NEWS-G:{\sc SEDINE}~\cite{Arnaud:2017bjh} and XENON1T (Midgal)~\cite{Aprile:2019jmx}.
  The lower yellow region is the neutrino floor for He:i-C$_4$H$_{10}$ (90\%:10\%).
  \proposalAcronym\ has the potential to explore significant regions of new parameter space beyond existing constraints and approaches the neutrino floor for DM masses around 0.6~GeV.}
\end{figure}

In Figure~\ref{fig:limits}, the projected sensitivity of \proposalAcronym\ in the parameter space of the DM mass and the spin-independent (SI) DM-nucleon cross-section is shown.
A running time of $300\;\si{days}$ and a 
flat background rate of $0.01\;\si{dru}$ was assumed. This is a conservative estimate of the background rate that is a factor $\sim2$ higher than the rate expected with a $2.5\,\si{\meter}$ water shield at Boulby (cf.\ Table~\ref{tab:summaryResults}). The $90\%$ Confidence Level (CL) exclusion limit was computed using a binned likelihood-ratio method with energy ranging between $14\;\si{\eV}$ and $1\;\si{\kilo\eV}$~\cite{Cowan:2010js}.
The projection assumes the Standard Halo Model with astrophysical parameters recommended in Ref.~\cite{Baxter:2021pqo}. 
A parameterisation of the Helm nuclear form factor from Ref.~\cite{Lewin:1995rx} was used, although $a=0.37$~fm, $s=0.99$~fm were used for helium and $a=0.47$~fm, $s=0.9$~fm for carbon, as these values provide a better fit to more recent calculations of the nuclear form factors (see e.g.,~\cite{Catena:2015uha, Gazda:2016mrp,Korber:2017ery,Andreoli:2018etf}). Although there are plans to measure the quenching factors in He:i-C$_{4}$H$_{10}$ gas (cf.\ Section~\ref{sec:quenching}), because of the lack of measurements at this time, SRIM~\cite{ZIEGLER20101818} was used to generate simulated quenching factors.
Finally, a COM-Poisson distribution with a Fano factor of $0.2$ was used to provide fluctuations in the primary ionisation~\cite{Agnese:2018gze} and a Polya distribution with $\theta = 0.12$ was used to generate fluctuations in the avalanche~\cite{Arnaud:2019nyp}.

Also in Figure~\ref{fig:limits}, projected sensitivity for {\sc S140} and {\sc ECuME} are shown, where the sensitivity curves have been computed using the optimal interval method~\cite{Yellin:2002xd} assuming background rates of $1.7\;\si{dru}$ and $0.3\;\si{dru}$, and exposures of $20\;\si{\kilo\gram \cdot days}$ and $200\;\si{\kilo\gram \cdot days}$, respectively~\cite{NEWS-G:2022kon}.
The solid lines in Figure~\ref{fig:limits} show existing constraints on the SI DM-nucleon cross-section so we see that \proposalAcronym\ explores extensive regions of new parameter space in the DM mass range from around $60\;\si{\mega\eV}$ to $5\;\si{\giga\eV}$.
The yellow shaded region in Figure~\ref{fig:limits} shows the parameter space where the coherent neutrino-nucleus interaction from solar neutrinos leads to a significant background (`the neutrino floor'). The He:i-C$_4$H$_{10}$ (90\%:10\%) neutrino floor has been calculated using the neutrino fluxes recommended in Ref.~\cite{Baxter:2021pqo} and the approach from Ref.~\cite{Billard:2013qya}, where the floor is the lower envelope of the background-free sensitivity curves for exposures that attain one neutrino event with threshold energies between $1\;\si{\eV}$ and $40\;\si{\kilo\eV}$. \proposalAcronym\ reaches the neutrino floor: we find that approximately two solar neutrino events are expected in a running time of 300 days.

\begin{figure}[t!] 
\includegraphics[width=0.95\columnwidth]{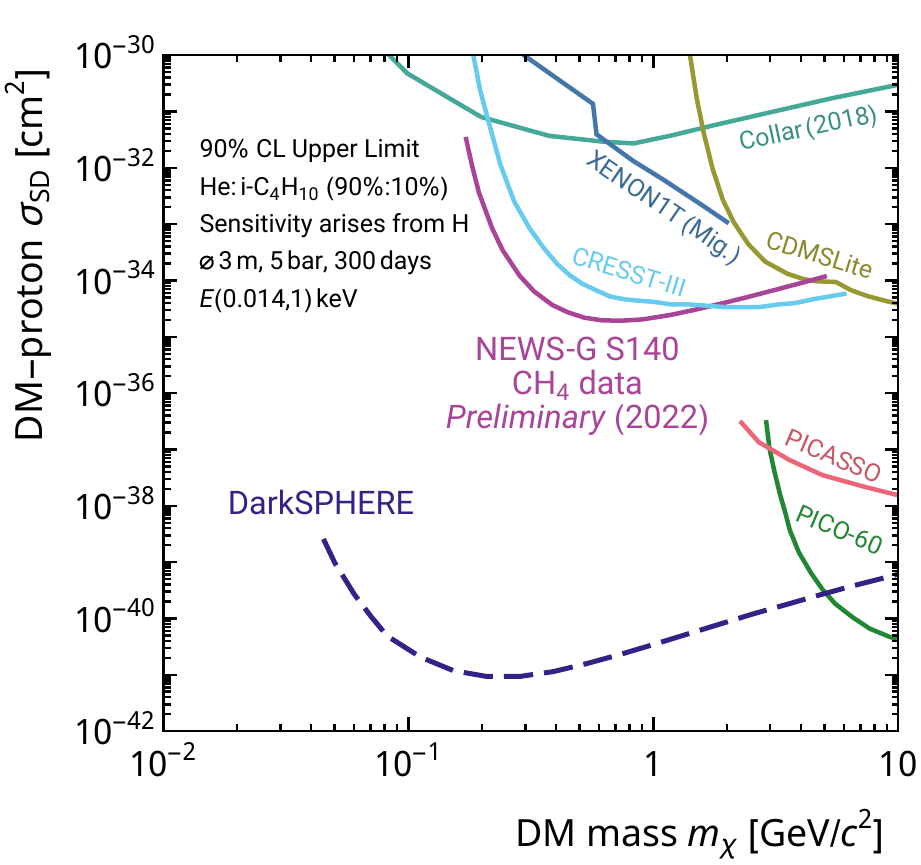}\\
\includegraphics[width=0.95\columnwidth]{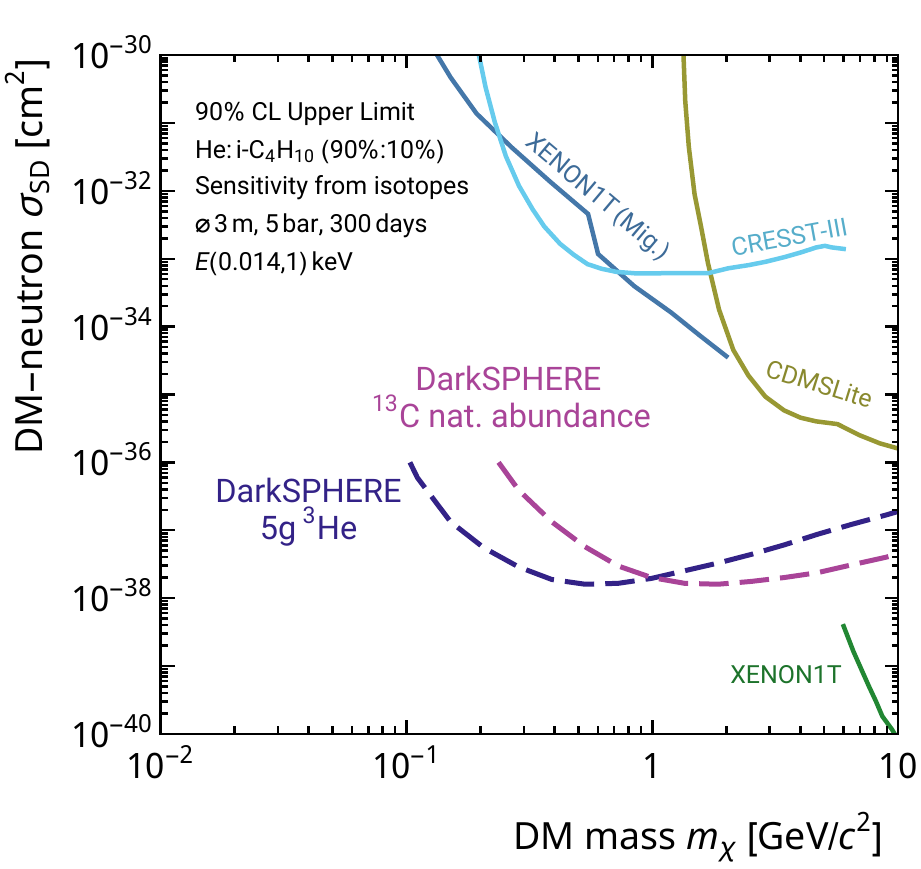}
  \caption{\label{fig:limitsSD}
    Sensitivity projections (dashed) for \proposalAcronym\ to the spin-dependent DM-nucleon cross-section. The upper panel shows the sensitivity when DM couples to the proton-spin. In  \proposalAcronym\, the sensitivity arises from hydrogen in the He:i-C$_{4}$H$_{10}$ gas. The lower panel shows the sensitivity when DM couples to the neutron-spin.
    In \proposalAcronym\, this sensitivity is achieved through the natural abundance of the $^{13}$C isotope, or by doping (at additional cost) with $^3$He.
  Solid lines show existing constraints from CDMSlite~\cite{SuperCDMS:2017nns}, Collar (2018)~\cite{Collar:2018ydf}, CRESST-III~\cite{CRESST:2022dtl}, PICASSO~\cite{Behnke:2016lsk}, PICO-60~\cite{PICO:2019vsc}, and 
  XENON1T (Midgal)~\cite{Aprile:2019jmx}.
  \proposalAcronym\ has the potential to explore significant regions of new parameter space beyond existing constraints.
}
\end{figure}

The sensitivity shown in Figure~\ref{fig:limits} is for the canonical SI DM-nucleus interaction that assumes equal couplings to protons and neutrons. 
However, the utilisation of hydrogen in the gas mixture, where the nucleus is a single spin-$1/2$ proton,  provides sensitivity to a wider range of effective field theory interactions that also depend on the nucleus spin~\cite{Fitzpatrick:2012ix}.
 For example, the upper panel of Figure~\ref{fig:limitsSD} shows the \proposalAcronym\
 sensitivity to the spin-dependent (SD) interaction with protons under the same assumptions used in the SI calculation. The total mass of hydrogen in the He:i-C$_{4}$H$_{10}$ (90\%:10\%) gas mixture at a pressure of $5\,\si{\bar}$ is $2.9\,\si{\kilo\gram}$. The shape of the \proposalAcronym\ projection 
 has a different shape compared to Figure~\ref{fig:limits} because the SD-proton sensitivity only comes from hydrogen, while the sensitivity in the SI case arises from hydrogen, helium and carbon. The upper panel of Figure~\ref{fig:limitsSD} also shows the preliminary results from the {\sc S140} experiment obtained with CH$_{4}$ gas (`S140 test data'). {\sc S140} provides the strongest constraint in the $0.2$-$2\,\si{\giga\eV}$ DM mass range~\cite{Blois} and demonstrates the advantage of running with a hydrogen target.
We again see that
 \proposalAcronym\ improves upon existing constraints by many orders of magnitude in the DM mass range from approximately $60\;\si{\mega\eV}$ to $5\;\si{\giga\eV}$.
 
Sub-dominant isotopes within the \proposalAcronym\ detector that have an unpaired neutron in the nucleus can also provide sensitivity to spin-dependent neutron interactions and therefore, the full panoply of spin-independent and spin-dependent effective field theory interactions can be tested with \proposalAcronym.
 Approximately~1.1\% of natural carbon contains the $^{13}$C isotope, which corresponds to a mass of approximately $14\;\si{\gram}$ in the He:i-C$_{4}$H$_{10}$ (90\%:10\%) mixture. The bottom panel of Figure~\ref{fig:limitsSD} presents the sensitivity of \proposalAcronym\ with the natural presence of $^{13}$C.
It may also be possible, albeit at additional cost, to dope the gas with a small amount of $^3$He and the resulting sensitivity from $5\,\si{\gram}$ of $^3$He is also shown. Although doping with $^3$He may be a logistical challenge, it would provide an additional handle to characterise the nature of the DM-nucleon interaction in the event of a discovery.
These projections use the averaged nuclear structure factor from values compiled in Ref.~\cite{Bednyakov:2004xq,Gazda:2016mrp}. 
As with the other nuclear recoil scenarios discussed, the lower panel of Figure~\ref{fig:limitsSD} shows that \proposalAcronym\ has the potential to test DM models with cross-sections orders of magnitude below the current constraints.

\begin{figure}[t!] 
  \includegraphics[width=0.95\columnwidth]{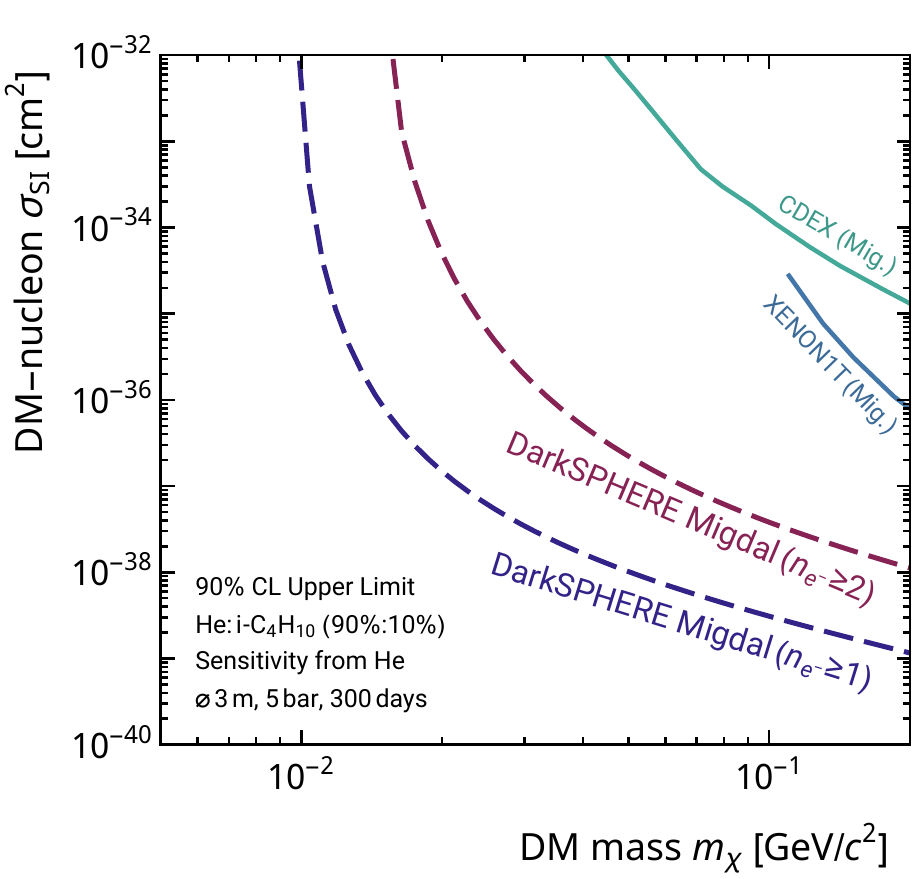}
  \caption{\label{fig:Migdallimits}
  Sensitivity projections (dashed) for \proposalAcronym\ to the spin-independent DM-nucleon cross-section that arises from the Migdal effect. 
  The two \proposalAcronym\ scenarios correspond to a single ($n_{e^-}\!\geq 1$) and double ($n_{e^-}\!\geq 2$) electron threshold.
  Solid lines show existing constraints from CDEX (Migdal)~\cite{CDEX:2021cll} and XENON1T (Midgal)~\cite{Aprile:2019jmx}.
  The Migdal effect gives \proposalAcronym\  the potential to explore significant regions of new parameter space below 0.1~GeV.}
\end{figure}

Next, we explore the sensitivity to the DM-nucleus interaction that can be achieved by exploiting the Migdal effect,
which accounts for the small probability that an electron can be emitted from an atom after the sudden perturbation of the nucleus through a scattering process with an electrically neutral projectile. Several collaborations have employed the Migdal effect to extend the sensitivity to lower DM masses (see e.g.,~\cite{LUX:2018akb, EDELWEISS:2019vjv,CDEX:2019hzn,XENON:2019zpr,SENSEI:2020dpa,CDEX:2021cll,EDELWEISS:2022ktt}). 

In Figure~\ref{fig:Migdallimits}, the projected sensitivity of \proposalAcronym\ in the parameter space of the DM mass and SI DM-nucleon cross-section is shown. Here, the focus is on a lower mass range than considered in Figure~\ref{fig:limits}.
A running time of 300 days and a flat background rate of 0.01~dru is again assumed, and the Migdal probability for helium is taken from Ref.~\cite{Cox:2022ekg}. Although the Migdal effect is expected to also apply to molecules including~i-C$_4$H$_{10}$, see e.g.~\cite{Lovesey:1982,Blanco:2022pkt}, probabilities for this molecule have not been calculated so we only include the Migdal effect from helium. We follow the simplified treatment described in Ref.~\cite{Hamaide:2021hlp} to model the detector response to the electron that has been ionised through the Migdal effect. The solid lines in Figure~\ref{fig:Migdallimits} show the leading constraints on the SI DM-nucleon cross-section derived using the Migdal effect. By exploiting the Midgal effect, we see that \proposalAcronym\ has the potential to explore  DM-nucleon interactions at even lower DM masses than shown in Figures~\ref{fig:limits} and~\ref{fig:limitsSD}.

Before leaving the topic of nuclear recoils, we discuss an approach that could allow \proposalAcronym\ to search for the DM-nucleus interaction from strongly-interacting, super-heavy DM with masses around the Planck scale.
 Such DM candidates could arise as fundamental states in theories of grand unification~\cite{Burdin:2014xma} and supersymmetry~\cite{Raby:1997pb}, or consist of macroscopic objects such as DM nuggets~\cite{Hardy:2014mqa}, soliton states~\cite{Ponton:2019hux} and primordial black holes~\cite{Lehmann:2019zgt}, to name a few. Owing to their large mass, the required number density to saturate the observed relic density is extremely low and becomes the limiting factor for direct detection. This means that the experimental sensitivity lies in the region where DM interacts strongly with nuclear matter, and invalidates the assumption that direct detection events only involve a single DM-nucleon scattering. Indeed, 
multiple scatterings are predicted for these states, both during its path through the experimental overburden and in the fiducial volume. This leads to drastically different signatures that require dedicated analysis and interpretation. In this regard, \proposalAcronym\ could benefit from the 60-anode ACHINOS read-out sensor which allows for track reconstruction, as discussed in Section~\ref{sec:sensorDev} in the context of background rejection, and would greatly aid a dedicated multiple-scattering search.

\begin{figure}[t!] 
  \includegraphics[width=0.95\columnwidth]{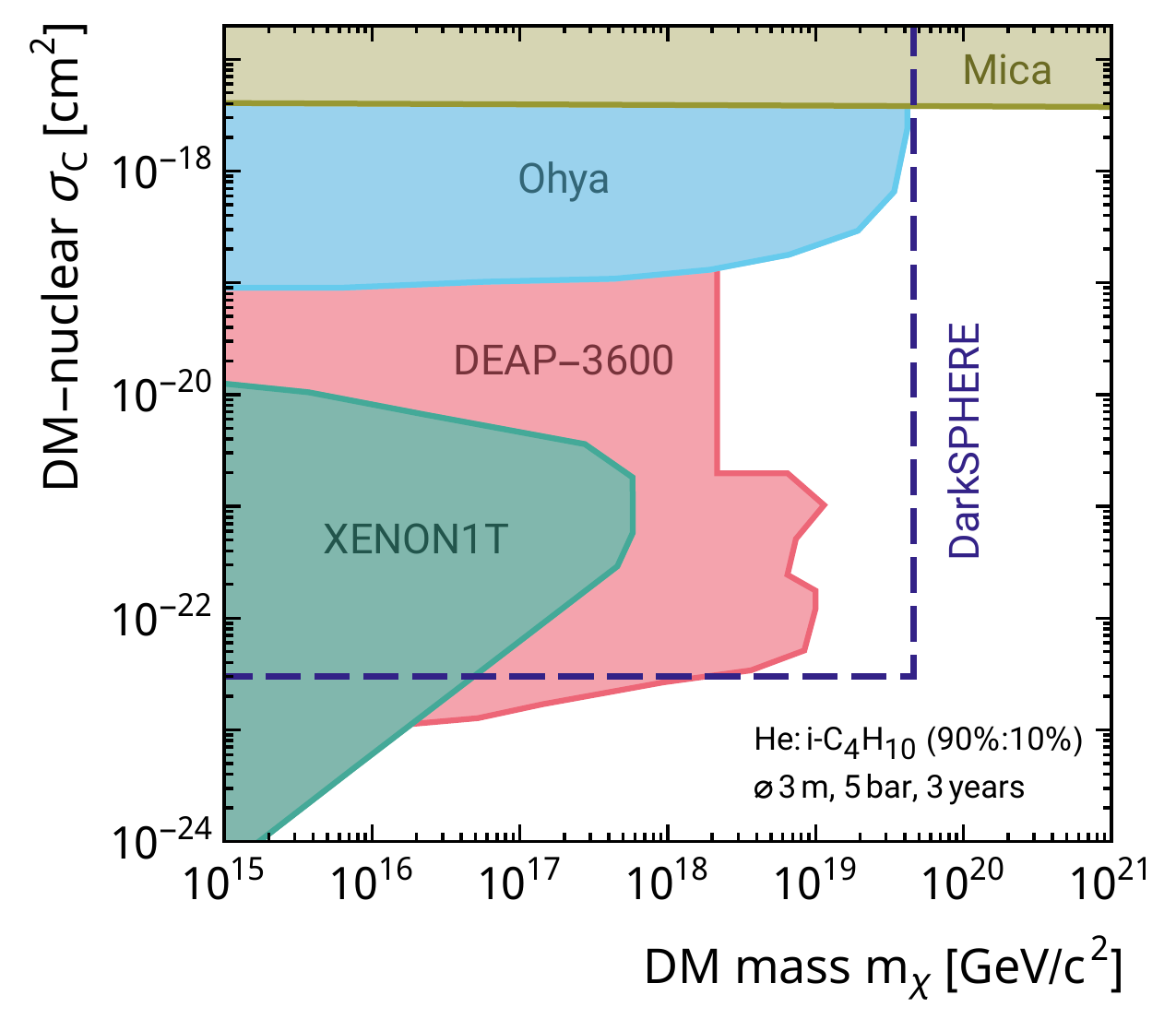}
  \caption{\label{fig:limitsHeavyDM}
Sensitivity projections (dashed) for \proposalAcronym\ to the contact DM-nuclear cross-section (which does not assume $A^4$ scaling, where $A$ is the atomic number). 
Shaded regions show existing constraints from
  `Model I' of the DEAP-3600 search (including the extrapolated regions)~\cite{Adhikari:2021fum},
  single scattering limits from XENON1T~\cite{Clark:2020mna}, etching plastic searches (Ohya)~\cite{Bhoonah:2020fys} and limits from ancient mica samples~\cite{Acevedo:2021tbl}.
  \proposalAcronym\ has the potential to explore unconstrained parameter space for DM masses around the Planck scale $(10^{19}~\mathrm{GeV})$.
}
\end{figure}

A consequence of the large DM-nuclear cross section is that, with such a high probability of scattering, the mass reach of a direct detection experiment scales as $m^{\text{max}}_{\chi}\propto A_{\text{det}}t_{\text{exp}}$, i.e., linearly with the cross-sectional area of the detector ($A_{\text{det}}$) and the exposure time ($t_{\text{exp}}$), and the DM-nuclear cross section sensitivity scales inversely with its diameter and the number density of target nuclei $\sigma_{C}\propto (L_{\text{det}}n_{\text{det}})^{-1}$~\cite{Bramante:2018qbc}.
With a diameter of $300~\!\mathrm{cm}$, \proposalAcronym\ would be one of the largest underground DM direct detection experiments. Comparing to the diameter of other direct detection experiments such as XENON1T ($100~\!\mathrm{cm}$~\cite{XENON:2017lvq}), LZ and XENONnT ($150~\!\mathrm{cm}$~\cite{LZ:2019sgr,XENON:2020kmp}) and DEAP-3600 ($170~\!\mathrm{cm}$~\cite{DEAP-3600:2017ker}) suggests that \proposalAcronym\ can offer a factor $3$--$10$ improvement in mass reach per exposure time. 

Figure~\ref{fig:limitsHeavyDM} shows the estimated reach in the DM mass -- contact DM-nuclear cross section ($m_{\chi},\sigma_C$) plane, assuming 3-years of exposure with a He:i-C$_{4}$H$_{10}$ (90\%:10\%) gas mixture at 5~bar. The projected \proposalAcronym\ sensitivity (horizontal and vertical lines) is estimated using the scaling formulae above.
The search for super-heavy DM can be carried out in parallel to the low-mass search as there are no online selections against tracks.
We contrast the sensitivity to existing limits from single scattering at \textsc{XENON1T}~\cite{Clark:2020mna}, etching plastic searches~\cite{Bhoonah:2020fys} and searches for evidence of DM scattering in ancient mica samples~\cite{Acevedo:2021tbl}. 
A dedicated search for multiply scattering massive particles by \textsc{DEAP-3600} is also shown~\cite{Adhikari:2021fum}. 
This estimate shows that \proposalAcronym\ has the potential to explore new parameter space for DM masses around the Planck scale (i.e.\ $m_{\chi}\sim 10^{19}~\!\mathrm{GeV}$) and motivates dedicated studies to obtain more precise projections.

\subsection{Electron ionisation from DM scattering}

\begin{figure}[t!] 
  \includegraphics[width=0.95\columnwidth]{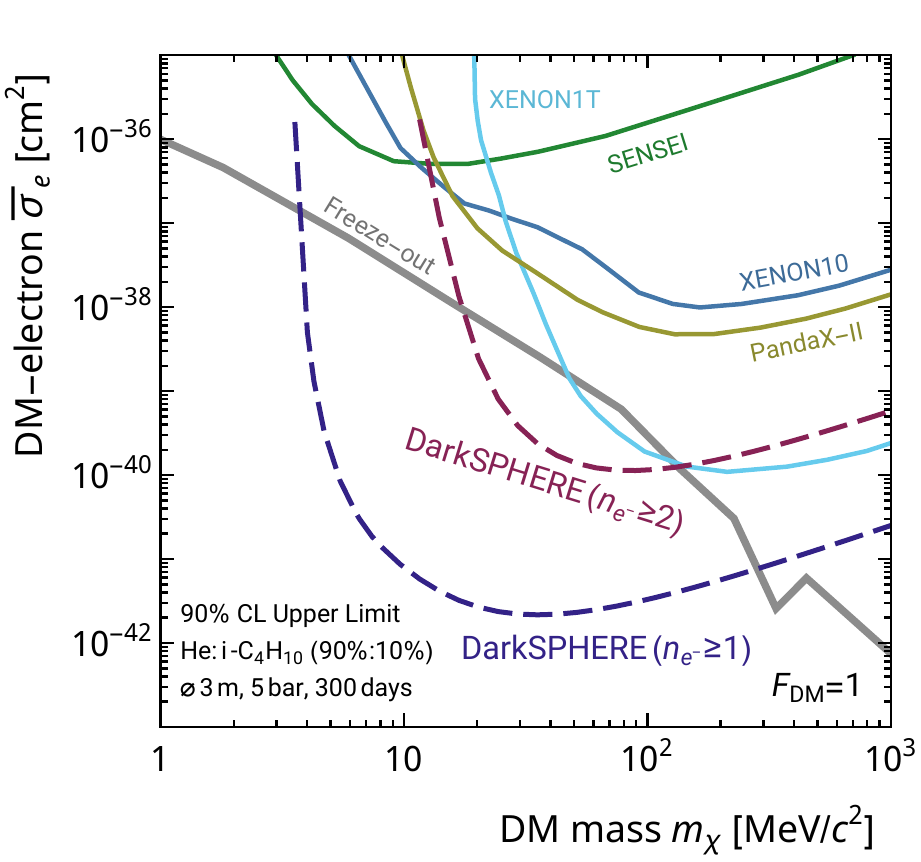}\\
  \includegraphics[width=0.95\columnwidth]{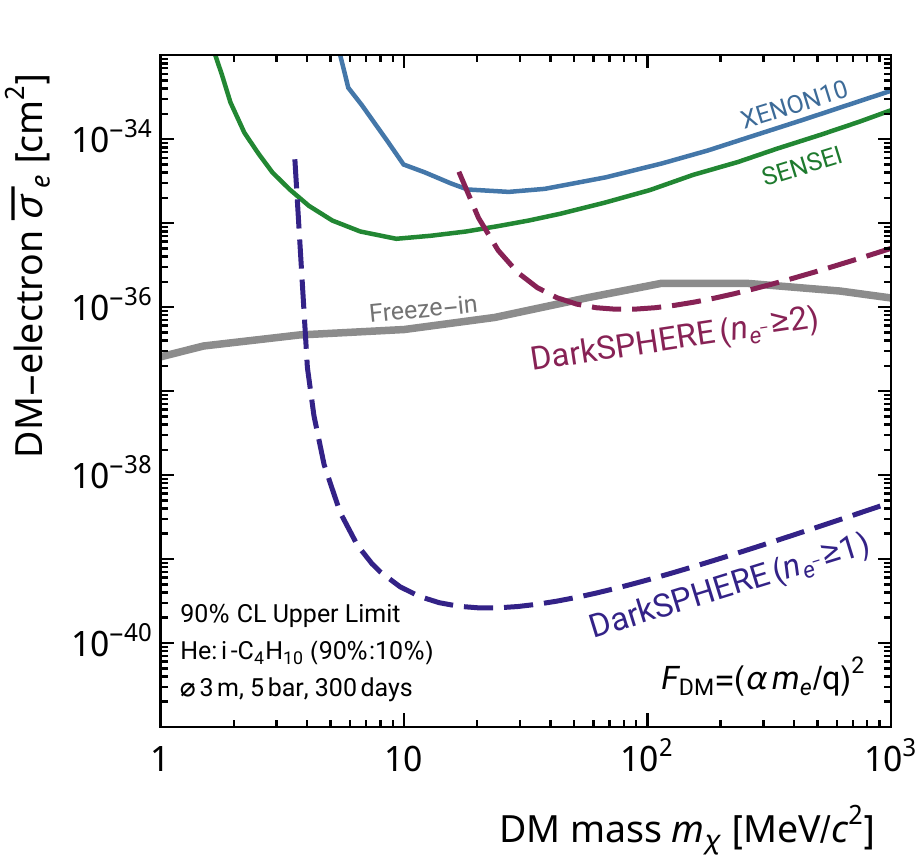}
  \caption{\label{fig:limits_electron}
  Sensitivity projections (dashed) for \proposalAcronym\ to the  DM-electron cross-section for two choices of the DM form factor: $F_{\rm{DM}}=1$ in the upper panel; and $F_{\rm{DM}}=(\alpha m_e/q)^2$ in the lower panel. The two \proposalAcronym\ scenarios correspond to a single ($n_{e^-}\!\geq 1$) and double ($n_{e^-}\!\geq 2$) electron threshold.
  Solid coloured lines show existing constraints from 
  SENSEI~\cite{SENSEI:2020dpa}, XENON10~\cite{Essig:2017kqs}, XENON1T~\cite{XENON:2019gfn} and PandaX-II~\cite{PandaX-II:2021nsg}. The grey lines show parameter space favoured by light DM benchmark models. \proposalAcronym\ has the potential to
  explore significant regions of unconstrained parameter space.
  }
\end{figure}

The single ionisation electron threshold of the spherical proportional counter, which is possible because of the small detector capacitance and high gain operation,
allows for the possibility of searching for DM-electron interactions~\cite{Hamaide:2021hlp}.\footnote{A single or few electron threshold also implies sensitivity to the absorption of keV-scale bosonic DM or keV-scale bosons produced in the Sun (see e.g.,~\cite{XENONCollaboration:2022kmb}). We reserve investigation of such scenarios for future work.}
The kinematics of DM-electron scattering in atoms and molecules mean that this search channel can probe DM candidates in the mass range from approximately 5~MeV to 1~GeV~\cite{Essig:2011nj}.
In \proposalAcronym, the signal induced by the DM interaction is an electron
that has been ionised from a helium atom or isobutane molecule in the He:i-C$_{4}$H$_{10}$ gas mixture, together with additional primary electrons generated as the ionised electron propagates through the gas.

In Figure~\ref{fig:limits_electron}, we show the projected sensitivity of \proposalAcronym\ in the parameter space of the DM mass and the DM-electron cross-section. The upper panel is for the DM form-factor $F_{\rm{DM}}=1$, corresponding to a DM-electron contact interaction, while the lower panel is for $F_{\rm{DM}}=(\alpha m_e/q)^2$, corresponding to an interaction by a light-mediator. Here, $\alpha$, $m_e$ and $q$ are the electromagnetic fine-structure constant, electron mass and momentum transfer respectively.
As with the nuclear recoil projections, we have assumed a running time of 300 days and a flat background of 0.01~dru, and again compute the 90\% CL exclusion limit using a binned likelihood ratio method with the astrophysical parameters recommended in Ref.~\cite{Baxter:2021pqo}. We use the dimensionless ionisation form factors for helium and isobutance provided in Ref.~\cite{Hamaide:2021hlp} to calculate the signal rate. For the detector response, we follow the simplified treatment described in Ref.~\cite{Hamaide:2021hlp} and use a $1 \,(28)$~eV threshold on the electron kinetic energy as a proxy for the single ($n_{e^-}\!\geq 1$) and double ($n_{e^-}\!\geq 2$) electron threshold. The $F_{\rm{DM}}=(\alpha m_e/q)^2$ scenario is particularly sensitive to the threshold so the sensitivity in this case decreases rapidly as the threshold is increased.

The solid coloured lines in Figure~\ref{fig:limits_electron} show the existing constraints on the DM-electron cross-section while the grey lines show the parameter space motivated by the freeze-out (top panel) and freeze-in (bottom panel) benchmark light DM models in Refs.~\cite{Essig:2011nj,Essig:2015cda}.
 We find that \proposalAcronym\ has the potential to significantly improve upon existing constraints under both the single and double electron threshold scenarios.
  \proposalAcronym\ is also able to probe the light DM benchmark models over a significant region of parameter space and complements other experimental proposals (e.g.,~\cite{Essig:2015cda,Bernstein:2020cpc, Castello-Mor:2020jhd,Aguilar-Arevalo:2022kqd}) by probing similar parameter space but with a different target medium and technology.
      
\subsection{Physics searches with a xenon-filled detector}

In addition to light DM searches, the simplicity of the spherical proportional counter's design, its low radioactive backgrounds, and background discrimination capabilities lend themselves to neutrinoless double $\beta$-decay searches. The Rare Decays with Radial Detector (R2D2) R\&D effort is exploring the use of spherical proportional counters for this end, with a goal to perform a neutrinoless double $\beta$-decay search with a tonne-scale $^{136}$Xe-filled detector~\cite{Meregaglia:2017nhx}. Recent effort has been made towards demonstrating the energy resolution~\cite{Bouet:2020lbp} and light-readout capabilities~\cite{Katsioulas:2021usd} that are required for R2D2. 
The low background and good energy resolution of a spherical proportional counter filled with $^{136}$Xe gas also make it a potential tool for supernova neutrino searches~\cite{Vergados:2005ny, Meregaglia:2017nhx}. For example, $^{136}$Xe at $5\,\si{\bar}$ in the \proposalAcronym\ detector would detect approximately $5\times (10\,\mathrm{kpc}/d_{\rm{SN}})^2$ events for a supernova explosion at a distance $d_{\rm{SN}}$, assuming a $27\,M_{\odot}$ progenitor~\cite{Lang:2016zhv}, while a negligible environmental background ($\lesssim10^{-3}$ events) would be expected during the short duration of the supernova burst.
As a multi-physics platform, \proposalAcronym\ could be used for such searches. 

\section{Summary and outlook}
\label{sec:summary}

\proposalAcronym\ offers the exciting potential for ground-breaking rare-event searches with a large, fully electroformed underground, spherical proportional counter. Previous experience of the NEWS-G collaboration, which is expanding with the physics exploitation of {\sc S140} and the construction of {\sc ECuME}, provides a solid foundation for the establishment of \proposalAcronym.

\proposalAcronym\ will improve upon {\sc S140} and {\sc ECuME} by increasing the diameter of the spherical proportional counter, operating the gas mixture under higher pressure, and by improving the shielding system to minimise the environmental background rate. These improvements necessitate 
R\&D for the scaling-up of the electroformation of spherical copper structures, 
an advanced 60-anode ACHINOS sensor with individual anode read-out, and a novel shield design. An integral part in the development of these advancements is the state-of-the-art simulation framework for spherical proportional counters that has been developed.

The project’s planned location is at the Boulby Underground Laboratory and will benefit from Boulby’s unique expertise with gaseous detectors. \proposalAcronym\ would further develop Boulby's expertise in hosting state-of-the-art astro-particle physics experiments by establishing underground electroforming capability, which is attractive for hosting future DM experiments.

The primary science goal of \proposalAcronym\ is to discover and characterise light DM-nucleon interactions for DM in the 0.05--10\,GeV mass range. We have demonstrated that \proposalAcronym\ approaches the neutrino floor for DM masses around 0.6 GeV and has the potential to probe extensive regions of new parameter space for spin-independent and spin-dependent interactions with both protons and neutrons.

The characteristics of a low-background spherical proportional counter equipped with a multi-anode ACHINOS sensor also make it suitable for other DM searches. As two examples, we have discussed super-heavy dark matter that leave tracks in the detector and MeV-scale DM that interacts with electrons. The technology also has the potential to search for neutrinoless double $\beta$-decay or neutrinos from a supernova explosion if filled with Xe-136 gas.

To conclude, \proposalAcronym\  leverages significant prior international expertise and investment in the use of electroformed spherical proportional counters as rare-event detectors, and is expected to offer significant advances well beyond the state-of-the-art in our understanding of sub-GeV DM candidates.

\section*{Acknowledgements} \label{sec:acknowledgements}
 This project has received funding from the European Union's Horizon 2020 research and innovation programme under the Marie Sk\l{}odowska-Curie grant agreement no 841261 (DarkSphere), no 895168 (neutronSPHERE), and no 101026519 (GaGARin). 
Support from the UK Research and Innovation -
Science and Technology Facilities Council (UKRI-STFC) is acknowledged: 
 CM is supported by grants No.\ ST/N004663/1, ST/T000759/1; 
 KM is also supported by grant No.\ ST/T000759/1; and 
 KN is supported by the University of Birmingham Particle Physics consolidated grant No.\ ST/S000860/1.
The project has received further support from UKRI-STFC  through grants No. ST/V006339/1 and
ST/W005611/1.
 LH is supported by the Cromwell Scholarship at King's College London.

 \bibliography{bibliography}
\end{document}